\documentclass[nofootinbib,aps,11pt]{revtex4-1}
\usepackage{graphicx}
\usepackage{hyperref}
\usepackage{cancel}
\usepackage{amssymb}
\usepackage{textcomp}
\usepackage{amsmath}
\usepackage{bm}
\usepackage{times}
\usepackage{epsfig}
\usepackage{color}
\newcommand{\GeV}{{\rm GeV}}

\begin{document}
\title{\LARGE Same-Sign Dilepton Signature in the Inert Doublet Model}
\bigskip
\author{Fa-Xin Yang}
\author{Zhi-Long Han}
\email{sps\_hanzl@ujn.edu.cn}
\author{Yi Jin}
\affiliation{
School of Physics and Technology, University of Jinan, Jinan, Shandong 250022, China}
\date{\today}

\begin{abstract}
	
 In this paper, we perform a detailed analysis on the same-sign dilepton signature in the inert doublet model. Focusing on the low dark matter mass region, we randomly scan the corresponding parameter space. Viable samples allowed by various constraints are obtained, among which twenty benchmark points are selected for further collider signature study. At hadron colliders, the same-sign dilepton signature is produced via $pp\to W^{\pm *}W^{\pm *}jj \to H^\pm H^\pm jj$ with the leptonic decay mode $ H^\pm \to HW^\pm (\to l^\pm \nu)$, where $H$ is the dark matter candidate. We investigate the testability of this signal at the high-luminosity LHC (HL-LHC) and the proposed 27 TeV high-energy LHC (HE-LHC). According to our simulation, the HL-LHC with $\mathcal{L}=3~ab^{-1}$ can hardly probe this signal. Meanwhile, for the HE-LHC with $\mathcal{L}=15~ab^{-1}$, it is promising to obtain a $5\sigma$ significance when $250~\text{GeV}\lesssim m_{H^\pm}-m_H\lesssim 300$ GeV with dark matter mass $m_H\sim 60$ or 71 GeV.

\end{abstract}

\maketitle

\section{Introduction}

Although the discovery of Higgs boson \cite{Aad:2012tfa,Chatrchyan:2012ufa} demonstrated the viability of the Standard Model (SM), there are convincing evidences of physics beyond SM, such as the origin of dark matter (DM) and tiny neutrino masses. Recent Plank data indicates that dark matter accounts for about 85\% of the total matter content in the universe \cite{Aghanim:2018eyx}. Among various candidates of particle DM, the Weakly Interacting Massive Particles (WIMPs) are the most popular recipes \cite{Bertone:2004pz,Arcadi:2017kky}, due to the fact that thermally produced WIMPs with weak-scale cross section can naturally lead to the observed DM relic density.

The Inert Doublet Model (IDM)~\cite{Deshpande:1977rw,Barbieri:2006dq,LopezHonorez:2006gr} is one of the simplest extensions of SM that provides DM candidate. This model introduces an inert Higgs doublet, which is odd under the unbroken $Z_2$ symmetry. There are four additional scalars as in the usual two Higgs doublet models \cite{Branco:2011iw}, i.e., neutral CP-even scalar ($H$), neutral CP-odd scalar ($A$), and charged scalar ($H^\pm$). The imposed unbroken $Z_2$ symmetry not only forbids Yukawa interactions of inert scalars with SM fermions, but also protects the lightest inert scalar being stable. In this paper, we consider the neutral CP-even scalar $H$ as DM candidate.
If we further introduce $Z_2$-odd right-hand neutrinos, the tiny neutrino masses can also be realized by the Scotogenic mechanism \cite{Ma:2006km,Han:2019diw,Han:2019lux,Wang:2019byi}. The phenomenology of the IDM has been extensively studied in Refs.~\cite{Gustafsson:2007pc,Cao:2007rm,Lundstrom:2008ai,Dolle:2009fn,Honorez:2010re,LopezHonorez:2010tb,Gustafsson:2012aj,Borah:2012pu,Swiezewska:2012eh,Osland:2013sla,Goudelis:2013uca,Modak:2015uda,Blinov:2015vma,Arhrib:2015hoa,Plascencia:2015xwa,Ilnicka:2015jba,Diaz:2015pyv,Kanemura:2016sos,Hashemi:2016wup,Belyaev:2016lok,Borah:2017dfn,Banerjee:2019luv,Jueid:2020rek,Abouabid:2020eik,Fabian:2020hny,Kalinowski:2020rmb,Banerjee:2021oxc,Banerjee:2021anv,Banerjee:2021xdp,Banerjee:2021hal}.

It is noticeable that the current positive evidences of DM all come from cosmological observations, which are based on the gravitational effects of DM. Therefore, the nature of DM is still an open question. To verify its nature, searches have been performed along three directions: direct detection, indirect detection, and collider signature. Despite the non-observation of direct detection signal which has already put stringent constraints on the parameter space of IDM \cite{Arhrib:2013ela,Belyaev:2016lok}, it is still appealing to extract positive indirect detection or collider signatures. For instance, low mass DM in IDM is possible to explain the Galactic center excess reported by Fermi-LAT \cite{Eiteneuer:2017hoh}. Meanwhile, a large parameter space of high mass DM in IDM is detectable at the Cherenkov Telescope Array \cite{Queiroz:2015utg,Garcia-Cely:2015khw}. As for collider searches, promising signatures are the dilepton \cite{Dolle:2009ft,Belanger:2015kga}, trilepton \cite{Miao:2010rg}, and teralepton channel\cite{Gustafsson:2012aj,Datta:2016nfz} at LHC. The vector boson fusion (VBF) channel $pp\to HHjj$ is also considered in Refs.~\cite{Dutta:2017lny,Dercks:2018wch}. Other promising collider signatures can be found in Refs.~\cite{Aoki:2013lhm,Arhrib:2014pva,Hashemi:2015swh,Belyaev:2018ext,Kalinowski:2018ylg,Kalinowski:2018kdn,Guo-He:2020nok}.

The same-sign pair production of charged Higgs bosons via vector boson fusion (VBF) in two Higgs doublet model was recently proposed by Ref.~\cite{Aiko:2019mww} to explore the nature of the Higgs potential, where two typical decay modes $H^\pm\to \tau \nu$ and $H^\pm \to tb$ are considered. The decay modes $H^\pm \to W^\pm A$ with $A\to b\bar{b}$ or $A\to \tau^+\tau^-$ are also studied in Ref.~\cite{Arhrib:2019ywg}. In this paper, we consider the decay mode $H^\pm \to W^\pm H$ with $H$ being the DM candidate, which leads to the same-sign dilepton signature $pp\to H^\pm H^\pm jj \to (W^\pm H)(W^\pm H)jj\to l^\pm l^\pm jj+\cancel{E}_T$. Notably, the well studied opposite-sign dilepton signature in IDM is only promising with compressed mass spectrum $\Delta m=m_A-m_H\in [40, 80]$ GeV \cite{Dolle:2009ft}. A distinct nature of the same-sign dilepton signature is that the production cross section will be enhanced when the mass splitting $\Delta m$ becomes larger \cite{Aiko:2019mww}. Meanwhile, the SM background of the same-sign dilepton signature \cite{Sirunyan:2017ret,Aaboud:2019nmv,Sirunyan:2020gyx,Sirunyan:2020gvn} is much smaller than the opposite-sign dilepton. Therefore, we expect that the same-sign dilepton signature might be promising for large $\Delta m$, which is complementary to the opposite-sign dilepton signature.

The paper is organized in the following way. In section \ref{SEC:TM}, we briefly review the inert doublet model. Focusing on the low mass region $m_H<100$ GeV, viable parameter space is explored by considering certain constraints. A detailed study of the same-sign dilepton signature is performed in section \ref{SEC:SDL}. Conclusion is presented in section \ref{SEC:CL}

\section{The Model}\label{SEC:TM}

In this paper, we consider the inert doublet model proposed in Ref.~\cite{Deshpande:1977rw,LopezHonorez:2006gr}.  In addition to the SM Higgs doublet $H_1$, an inert Higgs doublet $H_2$ is further introduced. The inert doublet $H_2$ is odd under an imposed $Z_2$ symmetry,  thus $H_2$ does not couple to SM fermions directly but to gauge bosons only. The $Z_2$ symmetry also ensures the stability of DM candidate. Provided the $Z_2$ symmetry is not broken spontaneously, then $H_2$ will not develop a vacuum expectation value (VEV). The Higgs doublets can be denoted as
\begin{align}
H_1=\left(
\begin{array}{c}
G^+\\
\frac{1}{\sqrt{2}}(v+h+i G^0)
\end{array}\right),\quad 
H_2 =
\left(
\begin{array}{c}
H^+\\
\frac{1}{\sqrt{2}}(H+ i A)
\end{array}\right),
\end{align}
where $G^\pm,G^0$ is the would-be Goldstone bosons, $v$ is the VEV of $H_1$, and $h$ is the SM Higgs boson.
The Higgs potential under the exact $Z_2$ symmetry is given by
\begin{eqnarray}
	V&=&\mu_1^2 H_1^\dag H_1+\mu_2^2 H_2^\dag H_2+\lambda_1 (H_1^\dag H_1)^2 + \lambda_2  (H_2^\dag H_2)^2 +\lambda_3  (H_1^\dag H_1) (H_2^\dag H_2) \\ \nonumber
	&&+ \lambda_4 (H_1^\dag H_2)(H_2^\dag H_1) +\frac{\lambda_5}{2} \left[ (H_1^\dag H_2)^2 +\text{h.c.} \right].
\end{eqnarray}
Here, all the free parameters are taken to be real. Due to the unbroken $Z_2$ symmetry, term as $\mu_{12}^2(H_1^\dag H_2 + H_2^\dag H_1)$ is forbidden. Therefore, $H_1$ and $H_2$ do not mix. After electroweak symmetry breaking, masses of scalars are given by
\begin{eqnarray}
	m_h^2 &=&-2\mu_1^2 = 2\lambda_1 v^2 \label{Eqn:mh}\\  
	m_H^2 &=& \mu_2^2 + \frac{1}{2}(\lambda_3+\lambda_4+\lambda_5)v^2 \label{Eqn:mH}\\  
	m_A^2 &=& \mu_2^2 + \frac{1}{2}(\lambda_3+\lambda_4-\lambda_5)v^2 \\  
	m_{H^\pm}^2 &=& \mu_2^2 +\frac{1}{2}\lambda_3 v^2
\end{eqnarray}
$H$ is taken to be the DM candidate in the following studies, which correspond to $\lambda_5<0$. For $A$ being DM candidate , one can simply make the replacement $\lambda_5\leftrightarrow -\lambda_5$. The parameters $\mu_1$ and $\lambda_1$ can be fixed by SM Higgs mass $m_h$ and VEV $v$. Then we are left with five free parameters, i.e., $\{\mu_2,\lambda_2,\lambda_3,\lambda_4,\lambda_5\}$. A more convenient set of parameters are $\{m_H,m_A,m_{H^\pm},\lambda_2,\lambda_L\}$, where  $\lambda_L=(\lambda_3+\lambda_4+\lambda_5)/2$ describes the Higgs-DM interaction $hHH$.

Extensively discussed in previous studies, the above parameter set is constrained by various theoretical and experimental bounds. Benchmark points, which satisfy all constraints, have been given in Ref.~\cite{Kalinowski:2018ylg}. Here, we briefly discuss the relevant constraints to the low mass region we adopted. More details can be found in Refs.~\cite{Belyaev:2016lok,Kalinowski:2018ylg}.  
\begin{itemize}
	\item {\bf Perturbativity:} The model is perturbative when the quartic couplings satisfy
	\begin{equation}\label{Eqn:PT}
		|\lambda_1,\lambda_2,\lambda_3,\lambda_4,\lambda_5|\leq 4\pi.
	\end{equation}
	\item {\bf Vacuum stability: } The stability of the Higgs potential at tree level is guaranteed by the bounded from below conditions
	\begin{equation}
		\lambda_1>0,\lambda_2>0,\lambda_3+2\sqrt{\lambda_1\lambda_2}>0,\lambda_3+\lambda_4-|\lambda_5|
		+2\sqrt{\lambda_1\lambda_2}>0.
	\end{equation}
	\item {\bf Global minimum:} In order to make sure the inert minimum being a local one, one needs \cite{Ginzburg:2010wa}
	\begin{equation}
		\frac{\mu_1^2}{\sqrt{\lambda_1}}\leq \frac{\mu_2^2}{\sqrt{\lambda_2}}.
	\end{equation}
	Using Eqn.~\eqref{Eqn:mh} and \eqref{Eqn:mH}, the above condition can be translated to 
	\begin{equation}
		\lambda_L\leq \frac{\sqrt{2\lambda_2}m_h v+2m_H^2}{2v^2}.
	\end{equation}
	\item {\bf Unitarity: } The unitarity of $S$-matrix from scattering processes among scalars and gauge bosons requires that the corresponding absolute eigenvalues of the scattering matrix should be less than $8\pi$ \cite{Arhrib:2012ia}.
	By requiring the unitarity conditions are valid up to about 10~TeV, the mass splittings are found to be in the region as \cite{Khan:2015ipa,Datta:2016nfz}
	\begin{equation}
		m_A-m_H\lesssim 300~\GeV,~m_{H^\pm}-m_H\lesssim 300~\GeV.
	\end{equation}
	\item {\bf Electroweak precision tests:} The inert Higgs doublet will contribute to the oblique $S$ and $T$ parameters. Analytic expressions can be found in Ref.~\cite{Belyaev:2016lok}. As for the experimental limits, we take the global fit result in Ref.~\cite{Baak:2014ora}
	\begin{equation}\label{Eqn:ST}
		S=0.06\pm 0.09,~T=0.01\pm 0.07,
	\end{equation}
		with correlation coefficient +0.91.
	\item {\bf Gauge boson widths:} The measurement of decay widths of gauge bosons $W^\pm$ and $Z$ indicate that masses of inert scalars should satisfy the following conditions
	\begin{equation}
		m_{A,H}+m_{H^\pm}>m_W,~m_A+m_H>m_Z,~2m_{H^\pm}>m_Z.
	\end{equation}
	Thus, decays of $W^\pm,Z$ to inert scalars are not kinetically open.
	\item {\bf Collider searches:} Searches for supersymmetric particles at LEP via dijet or dilepton signal have excluded the following mass region \cite{Lundstrom:2008ai}
	\begin{equation}\label{Eqn:LEP}
		m_A\leq 100~\GeV,~m_H\leq 80~\GeV,~m_A-m_H\geq 8~\GeV,
	\end{equation}
	when the above conditions are satisfied simultaneously. Meanwhile, searches for chargino have set a lower limit on the charged scalar \cite{Pierce:2007ut}
	\begin{equation}
		m_{H^\pm}\geq 70 ~\GeV
	\end{equation}
	\item {\bf SM Higgs data:} The Higgs invisible decay channel gets additional contribution when DM is light enough, i.e., $m_H<m_h/2$. The current experimental limit on the branching ratio of Higgs invisible decay is \cite{Khachatryan:2016whc}
	\begin{equation}
		\text{BR}(h\to \text{invisible})<0.24.
	\end{equation}
	The charged scalar $H^\pm$ will also impact the Higgs to diphoton channel via one loop contribution \cite{Swiezewska:2012eh}. The experimental signal strength of diphoton is \cite{Khachatryan:2016vau}
	\begin{equation}
		\mu_{\gamma\gamma}=1.14^{+0.38}_{-0.36}.
	\end{equation}
	\item {\bf Relic density:} The DM relic density observed by the Planck experiment is \cite{Aghanim:2018eyx}
	\begin{equation}\label{Eqn:RD}
		\Omega h^2 = 0.1200\pm 0.0012.
	\end{equation}
	We require that the theoretical DM relic density of $H$ is within $3\sigma$ range of the observed value. {\bf MicrOmegas} \cite{Barducci:2016pcb} is used to calculate the relic density.
	\item {\bf Direct detection:} In this paper, we take the direct detection limit on the spin-independent cross section from the XENON1T experiment \cite{Aprile:2018dbl}, which is the most stringent one at present.
\end{itemize}

Focus on the low mass region, we randomly scan the parameter space in the following regions
\begin{align}
	m_H\in[50,80]~\GeV, ~\lambda_L\in[-0.04,0.04],~\lambda_2\in[0,1]\\ \nonumber 
	m_A-m_H\in[0,300]~\GeV,~m_{H^\pm}-m_H\in[0,300]~\GeV
\end{align}

\begin{figure}
	\begin{center}
		\includegraphics[width=0.46\linewidth]{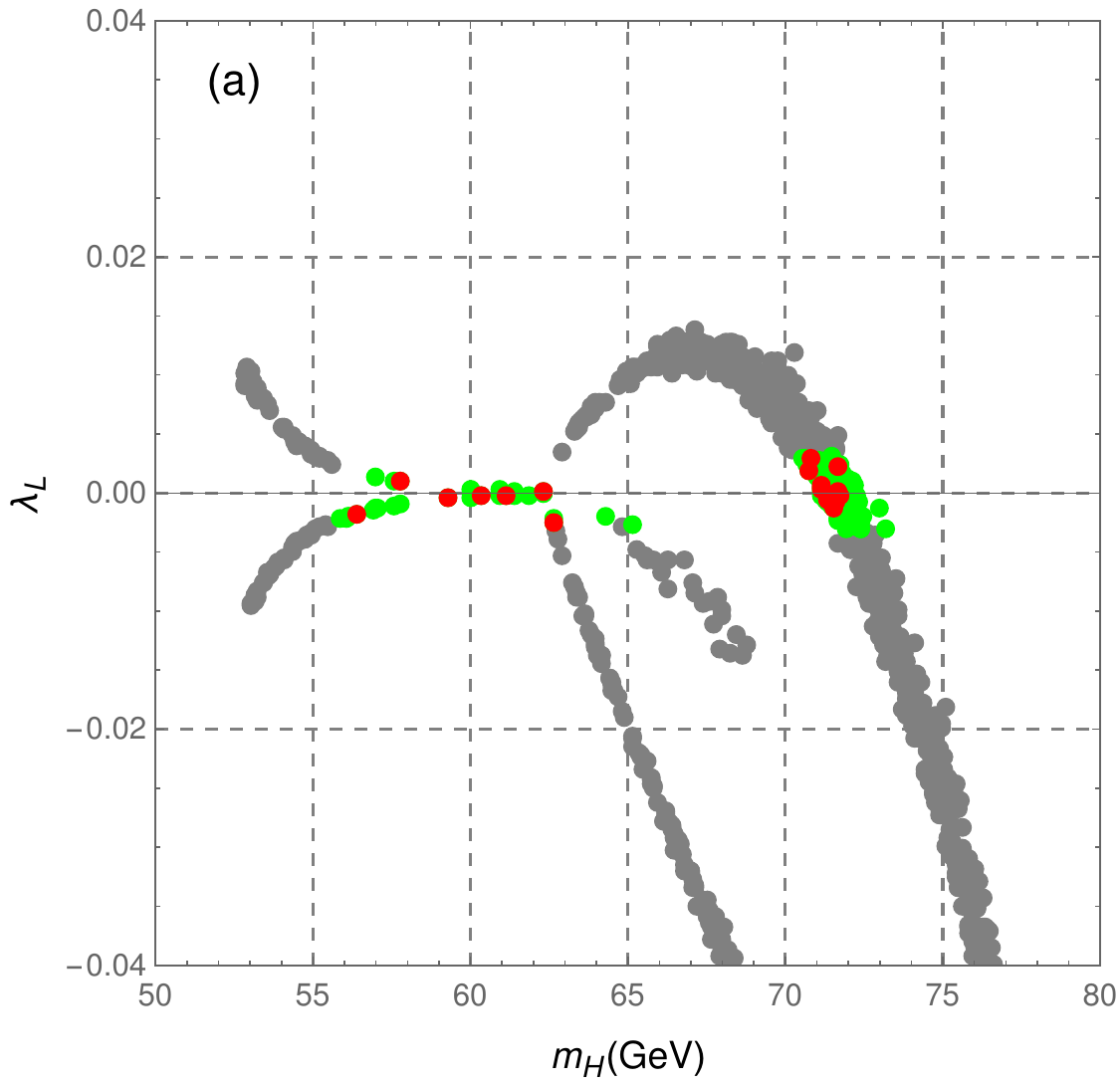}
		\includegraphics[width=0.45\linewidth]{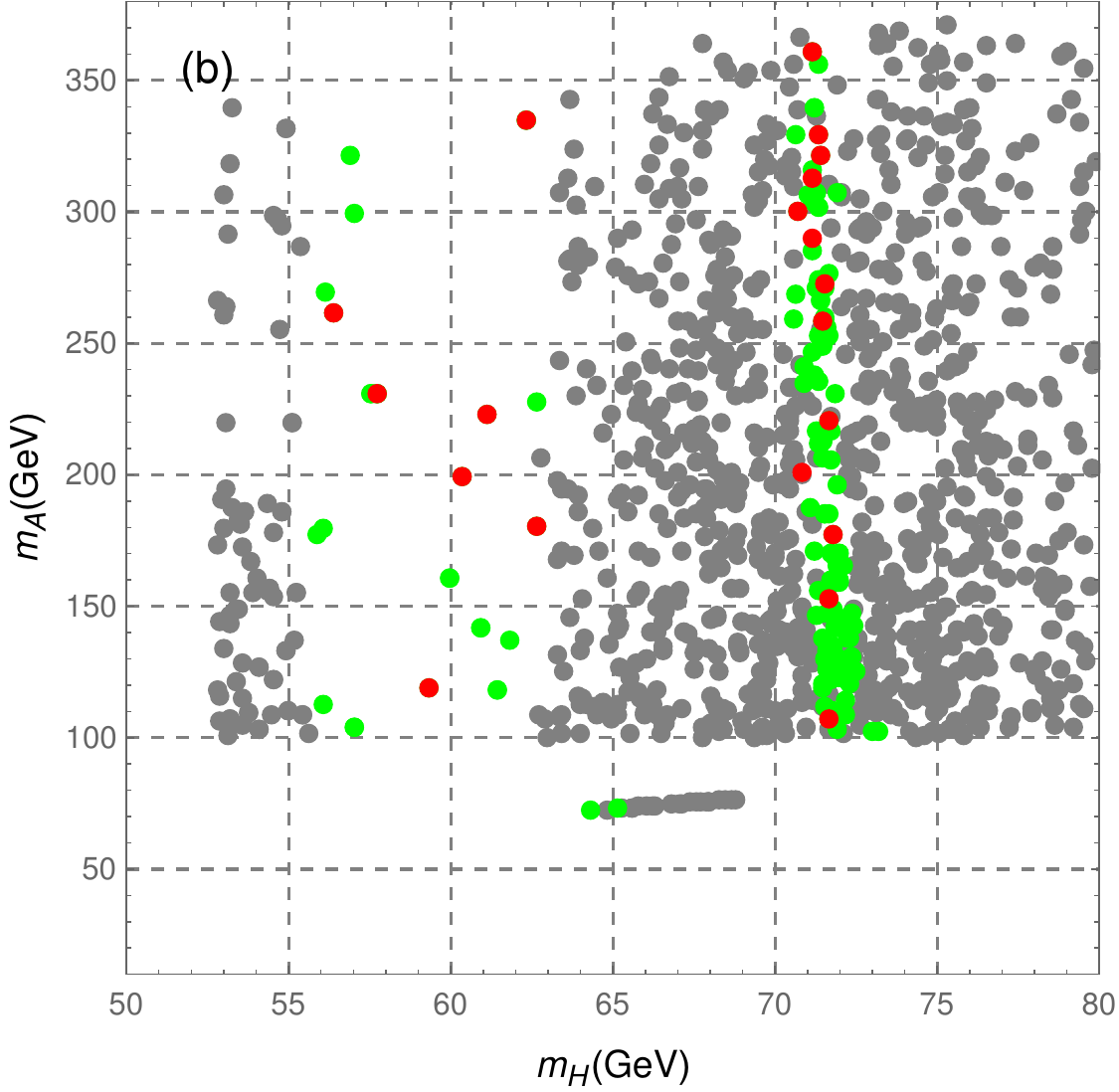}
		\\~
		\includegraphics[width=0.445\linewidth]{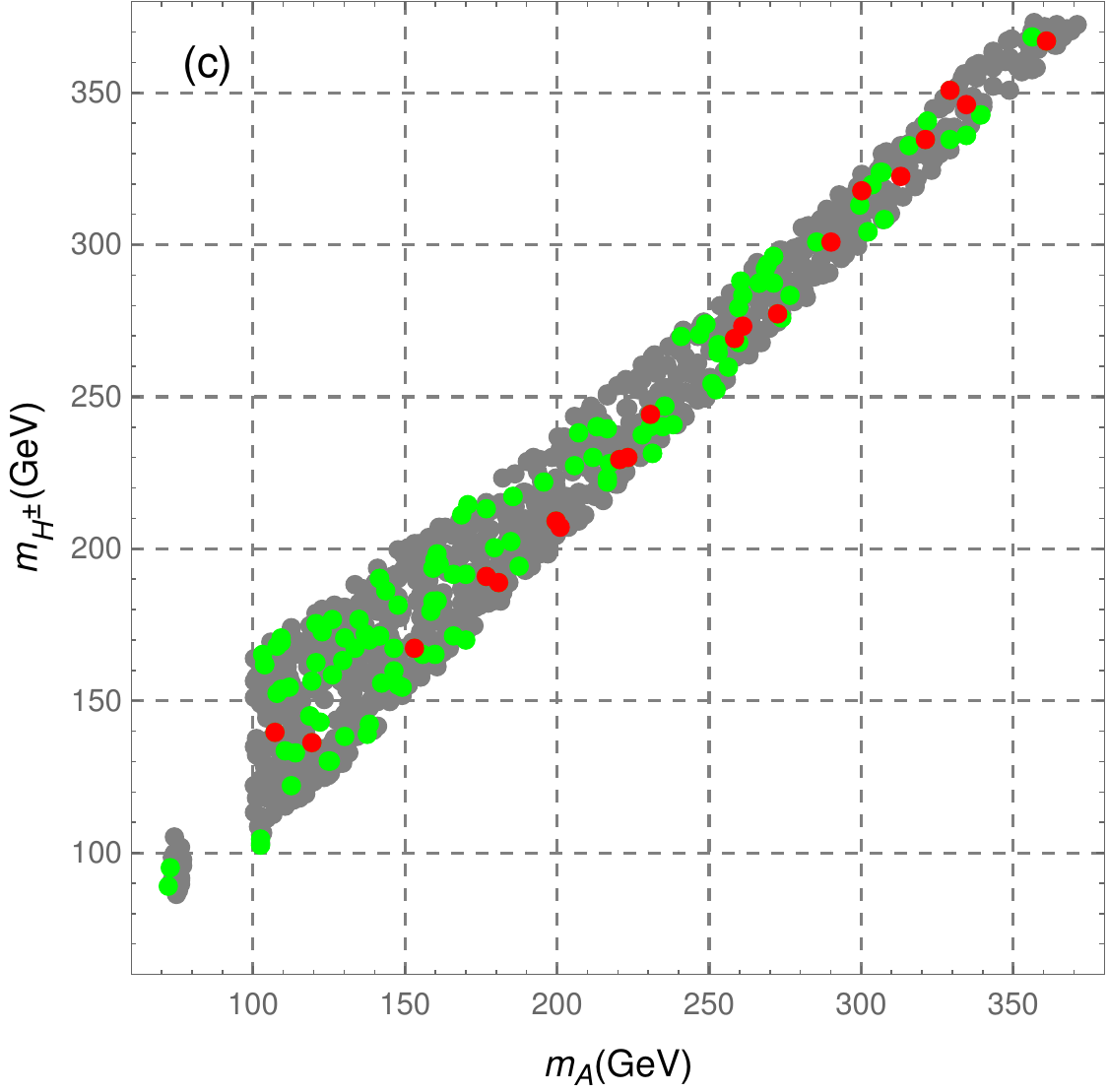}
		\includegraphics[width=0.45\linewidth]{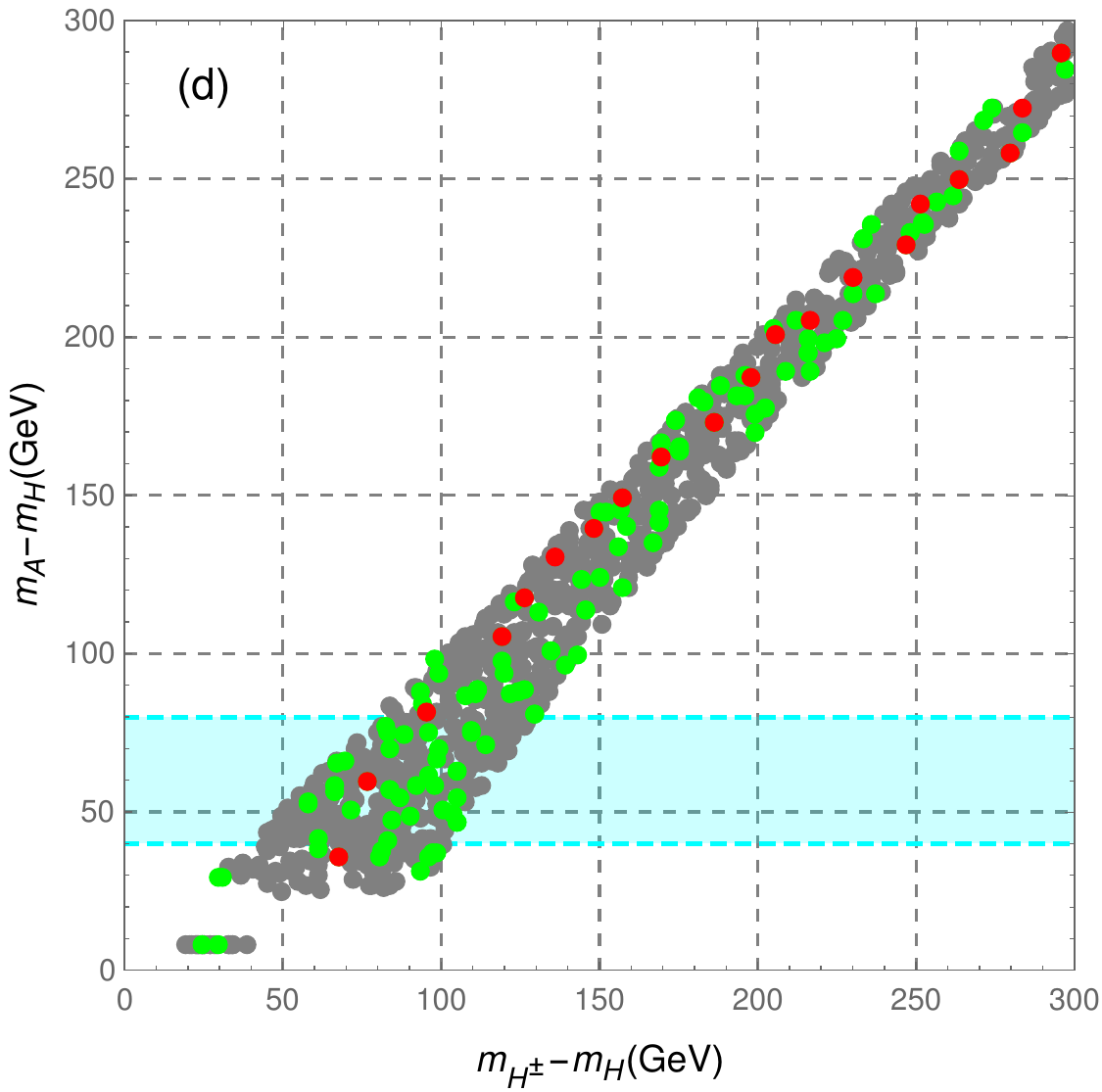}
	\end{center}
	\caption{Scanned results of the low mass region. Distribution of samples in the $(m_H,\lambda_L)$ plane (panel a), $(m_A,m_H)$ plane (panel b), $(m_{H^\pm},m_A)$ plane (panel c) and $(m_{H^\pm}-m_H,m_A-m_H)$ plane (panel d). All the samples satisfy constraints from Eqn.~\eqref{Eqn:PT} to \eqref{Eqn:RD}. The gray points are further excluded by the XENON1T result \cite{Aprile:2018dbl}. The green and red points are allowed by all constraints. The red points, which have been listed in Table \ref{Tab:BP} , are the benchmark points selected for following same-sign dilepton signature. The light blue band in panel d corresponds to the promising region of the opposite-sign dilepton signature \cite{Dolle:2009ft}.
		\label{Fig:Scan}}
\end{figure}

Scanned results are shown in Fig.~\ref{Fig:Scan}. The requirement of relic density within $3\sigma$ range together with the direct detection limit from XENON1T strictly constrain the parameter space. From  Fig.~\ref{Fig:Scan} (a) and (b), it is clear that the allowed samples of our scan fall into three separated regions. One is the Higgs resonance region around $m_H\lesssim m_h/2$. Another one is the vector boson annihilation region around $m_H\sim71.5$ GeV, where the dominant annihilation channel is $HH\to VV (V=Z,W)$. The mass region  $63~\GeV\lesssim m_H\lesssim 71$ GeV with $m_A>100$ GeV is now excluded by XENON1T. The third one is  the narrow coannihilation region $m_A-m_H\sim 8$ GeV. Since degenerate $m_A$ and $m_H$ will lead to vanishing same-sign charged Higgs pair at LHC \cite{Aiko:2019mww}, we will not consider such coannihilation region in the following. In Fig.~\ref{Fig:Scan} (c), results in the $(m_A,m_{H^\pm})$ plane are also shown. All the survived points satisfy $m_A\lesssim m_{H^\pm}$, mainly due to constraints from $S$ and $T$ parameters in Eqn.~\ref{Eqn:ST}. The mass gap between $80~\GeV\lesssim m_A<100$ GeV corresponds to the excluded region of LEP in Eqn.~\eqref{Eqn:LEP}. Because the same-sign dilepton signature is sensitive to the mass splitting $\Delta m=m_A-m_H$, corresponding results are also depicted in Fig.~\ref{Fig:Scan} (d).

\begin{table}
	\begin{tabular}{|c|c|c|c|c|c|c||c|c|}
		\hline\hline
	   ~ No.~~ & $m_H$(GeV) & $m_A$(GeV)  & $m_{H^\pm}$(GeV)  & ~$\lambda_2$~ & ~$\lambda_L$~ & ~$\Omega h^2$~ & $\sigma$ @14TeV (fb) & $\sigma$ @27TeV (fb)
	    \\ \hline
	    BP1  & 71.69 & 107.5 & 139.6 & 0.4097   & 0.002203  & ~0.1210~ & 0.054 & 0.160
	    \\ \hline
	    BP2  & 59.30 & 119.1 & 136.3 &~0.09806~ &~-0.0004655~& 0.1213  & 0.154 & 0.451
	     \\ \hline
	    BP3  & 71.67 & 152.9 & 167.0 & 0.1750   & ~0.0001029 & 0.1233  & 0.214 & 0.657 
	     \\ \hline
	    BP4  & 71.76 & 177.0 & 190.9 & 0.3855   & -0.0002066 & 0.1180  & 0.285 & 0.914
	     \\ \hline
	    BP5  & 62.64 & 180.5 & 189.1 & 0.7473   & -0.002478  & 0.1177  & 0.355 & 1.139
	     \\ \hline
	    BP6  & 70.82 & 201.1 & 206.8 & 0.8602   & 0.002879   & 0.1233  & 0.373 & 1.232
	     \\ \hline
	    BP7  & 60.37 & 199.7 & 208.8 & 0.6200   & -0.0002771 & 0.1210  & 0.409 & 1.351
	     \\ \hline
	    BP8  & 71.63 & 220.8 & 229.1 & 0.5264   & -0.0007215 & 0.1193  & 0.399 & 1.362
	     \\ \hline
	    BP9  & 61.12 & 223.2 & 230.3 & 0.4692   & -0.0002002 & 0.1227  & 0.454 & 1.553
	     \\ \hline
	    BP10 & 57.76 & 230.7 & 244.3 & 0.9192   & ~0.0009435 & 0.1185  & 0.454 & 1.578
	     \\ \hline
	    BP11 & 71.44 & 258.6 & 269.0 & 0.6848   & -0.0007471 & 0.1214  & 0.446 & 1.616 
	     \\ \hline
	    BP12 & 71.55 & 272.6 & 277.1 &~0.00294  & -0.001236  & 0.1205  & 0.483 & 1.765
	     \\ \hline
	    BP13 & 56.40 & 261.4 & 273.1 & 0.5082   & -0.001733  & 0.1191  & 0.495 & 1.799
	     \\ \hline
	    BP14 & 71.17 & 290.1 & 301.2 & 0.5216   & ~0.0006213 & 0.1200  & 0.467 & 1.788
	     \\ \hline
	    BP15 & 70.72 & 299.9 & 317.8 & 0.7495   & ~0.001944  & 0.1235  &  0.451 & 1.755
	     \\ \hline
	    BP16 & 71.12 & 312.9 & 322.7 &~0.04812  & ~0.0002456 & 0.1221  & 0.482 & 1.892
	     \\ \hline
	    BP17 & 71.39 & 321.4 & 334.9 & 0.7437   & -0.0001886 & 0.1172  & 0.468 & 1.883
	     \\ \hline
	    BP18 & 71.31 & 329.1 & 350.8 & 0.1182   & -0.0005298 & 0.1204  & 0.441 & 1.813
	     \\ \hline
	    BP19 & 62.32 & 334.6 & 346.0 & 0.2196   & ~0.0001064 & 0.1180  & 0.498 & 2.037
	     \\ \hline
	    BP20 & 71.14 & 360.8 & 366.8 & 0.1079   & ~0.0005207 & 0.1192  & 0.495 & 2.087
		\\\hline \hline
	\end{tabular}
	\caption{Benchmark points (BP) for the same-sign dilepton signature. Here, $\sigma$ denotes the cross section of $pp\to H^\pm H^\pm jj$ with preselection cuts in Eqn.\eqref{Eqn:Cut1}.\label{Tab:BP}}
\end{table}

Based on the above scanned results, we have selected 20 BPs (red ones in Fig.~\ref{Fig:Scan}) for the following study. Detailed information on these BP can be found in Table \ref{Tab:BP}. Different from Ref.~\cite{Kalinowski:2018ylg}, we have selected more BPs with  $\Delta m >150$ GeV. For BP1 to BP10, they could also be probed at the 380 GeV CLIC with 1 ab$^{-1}$ data, meanwhile the rest ten BPs are within the reach of 1.5 TeV CLIC with 2.5 ab$^{-1}$ data \cite{Kalinowski:2018kdn}.

\section{Same-Sign Dilepton Signature}\label{SEC:SDL}

\begin{figure}
	\begin{center}
		\includegraphics[width=0.45\linewidth]{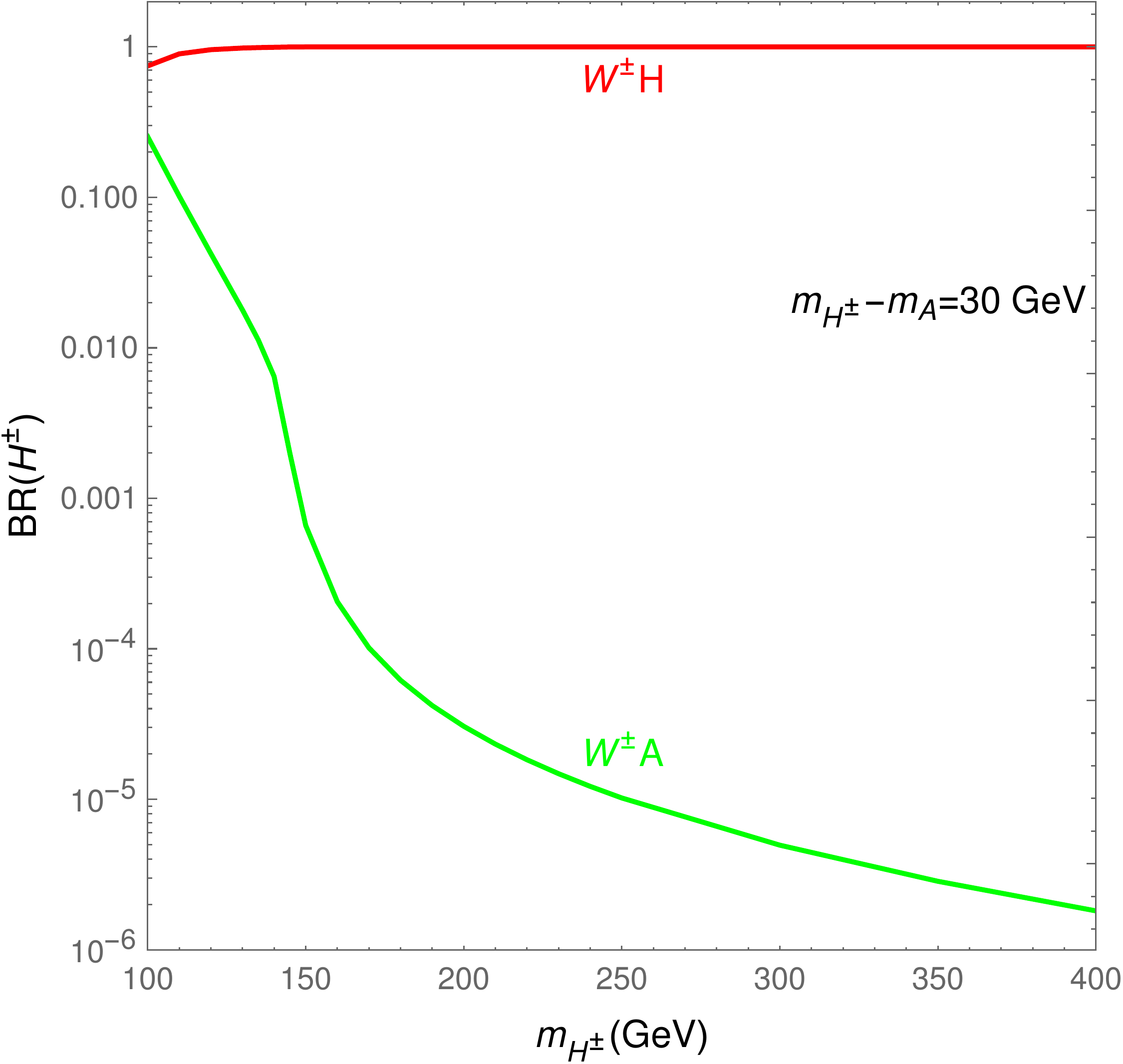}
		\includegraphics[width=0.45\linewidth]{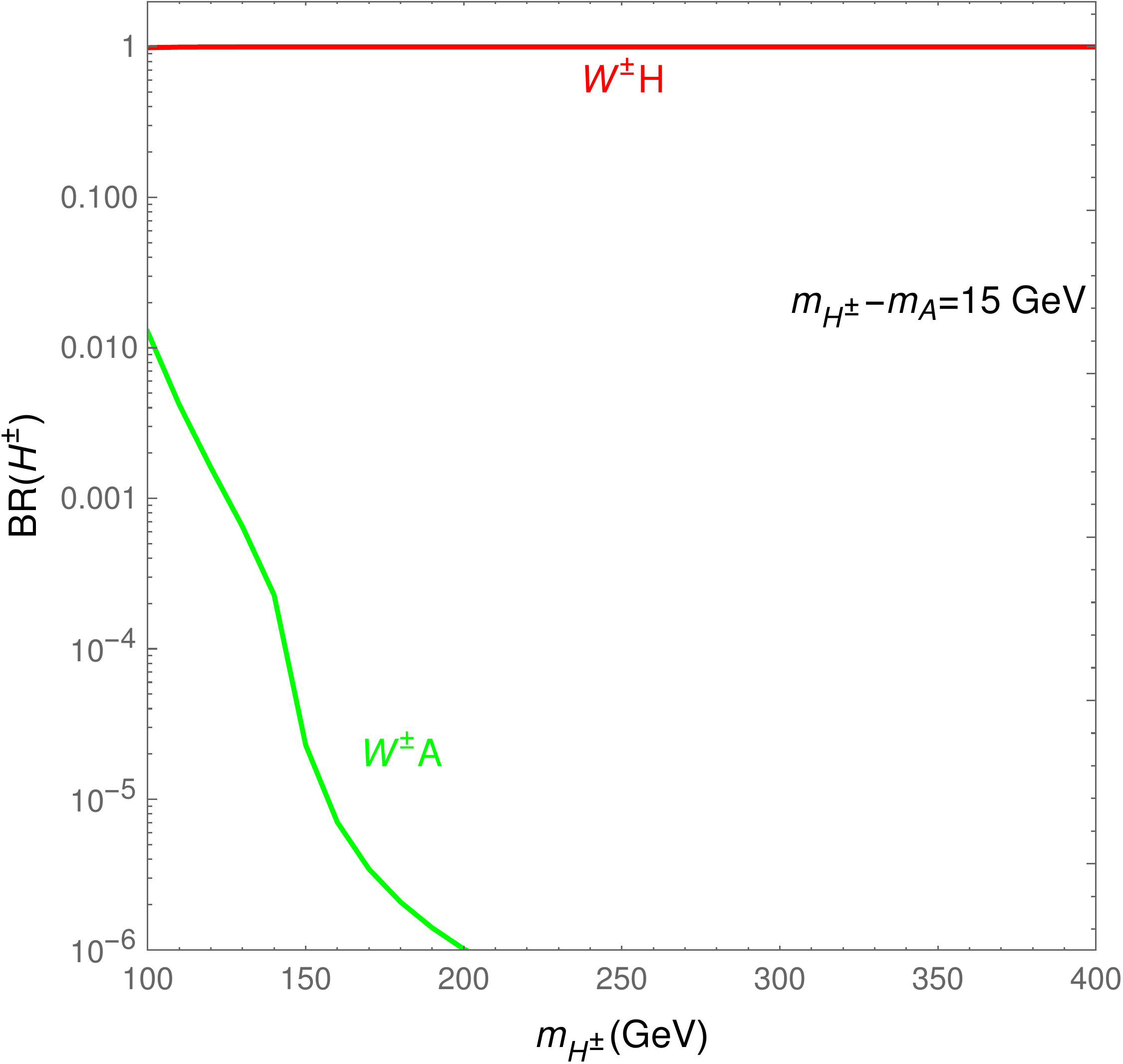}
	\end{center}
	\caption{Branching ratio of the charged scalar $H^\pm$ for $m_{H^\pm}-m_A=30 $ GeV (left panel) and $m_{H^\pm}-m_A=15 $ GeV (right panel), where $m_H$ is fixed to be 62 GeV in both cases. The package {\bf 2HDMC }\cite{Eriksson:2009ws} is used for calculating these branching ratios.
		\label{Fig:BRHP}}
\end{figure}

Before discussing the same-sign dilepton signature, we first consider the branching ratio of the charged scalar $H^\pm$. There are two possible decay modes of $H^\pm$ in the IDM. One is $H^\pm\to W^\pm H$, and the other is $H^\pm \to W^\pm A$. For the special scenario $m_H\sim m_A<m_{H^\pm}$, one would have $\text{BR}(H^\pm\to W^\pm H)\approx \text{BR}(H^\pm\to W^\pm A)\approx 0.5$. However, the precise measurement of $S$ and $T$ parameters requires $m_H<m_A\lesssim m_{H^\pm}$, which leads to a phase space suppression of the  $H^\pm \to W^\pm A$ mode. In Fig.~\ref{Fig:BRHP}, we illustrate the branching ratio of $H^\pm$. For $m_{H^\pm}-m_A=30 (15)$ GeV, we have $\text{BR}(H^\pm\to W^\pm A)<0.01$, i.e., $\text{BR}(H^\pm\to W^\pm H)>0.99$ when $m_{H^\pm} > 135 (100)$ GeV. That is to say, $H^\pm\to W^\pm H$ is always the dominant decay mode (approximate to one) for the BPs in Table \ref{Tab:BP}.

\begin{figure}
	\begin{center}
		\includegraphics[width=0.48\linewidth]{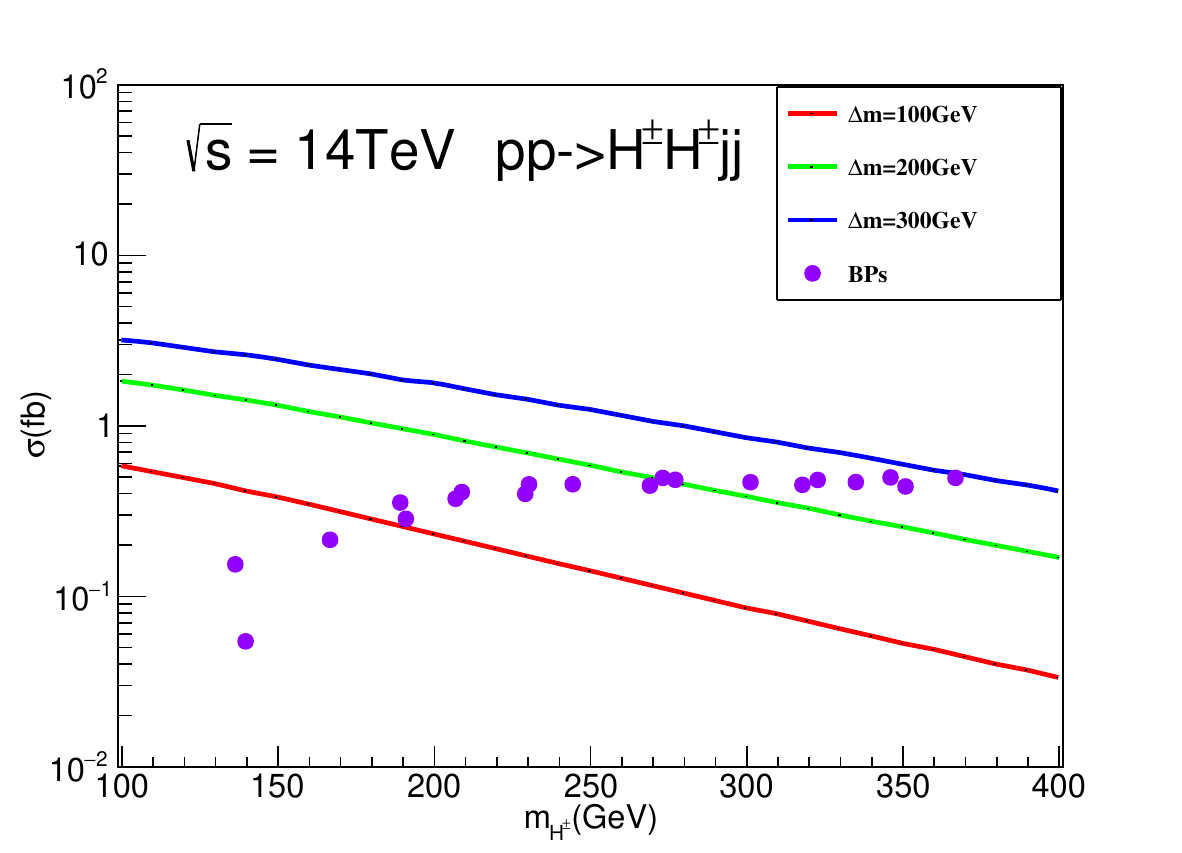}
		\includegraphics[width=0.48\linewidth]{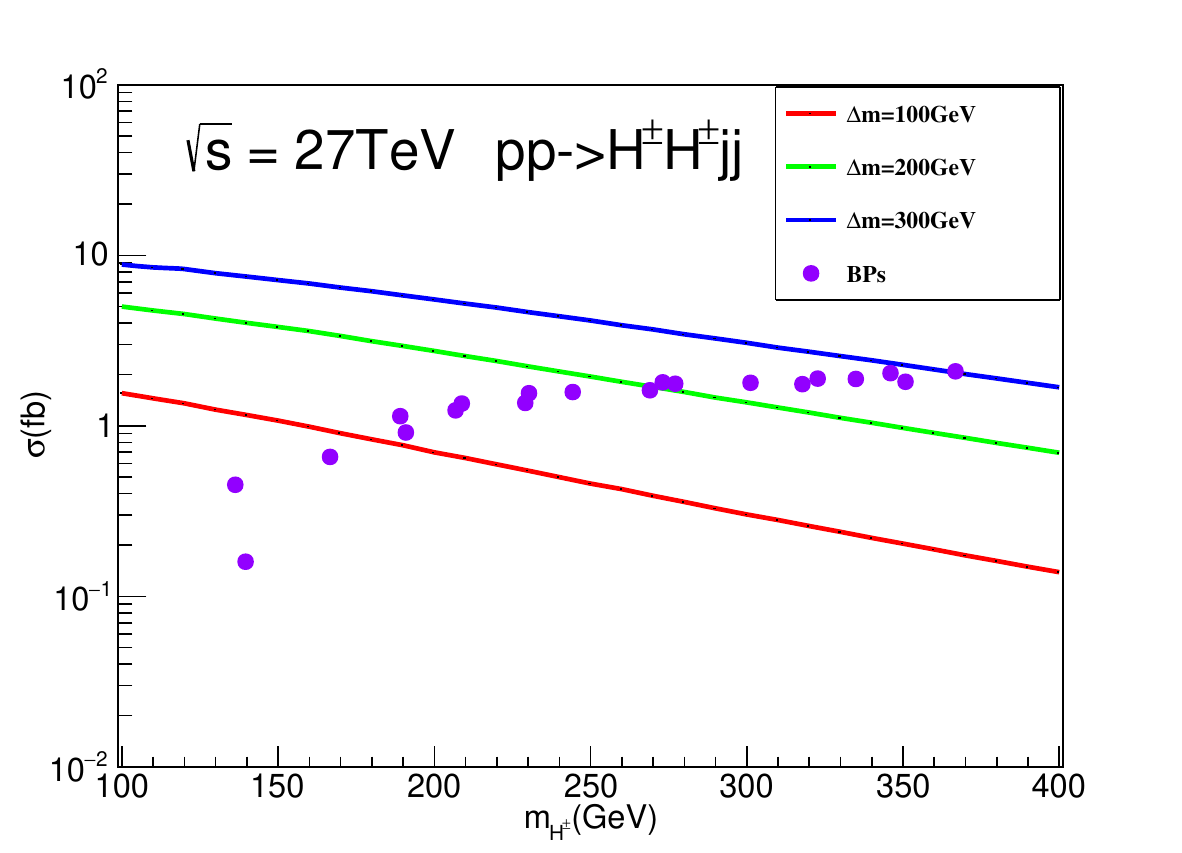}
	\end{center}
	\caption{Production cross section of process $pp\to H^\pm H^\pm jj$ at the  $\sqrt{s}=14$ TeV HL-LHC (left panel) and the $\sqrt{s}=27$ TeV HE-LHC (right panel) as a function of $m_{H^{\pm}}$  with $\Delta{m}=$ 100GeV, 200GeV, 300GeV, respectively. Here, we also fix $m_H=62$ GeV. The cross section of the BPs listed in Table \ref{Tab:BP} are also shown. Note that the preselection cuts in Eqn.\eqref{Eqn:Cut1} are already applied.}
	\label{Fig:CS}
\end{figure}

An essential feasibility of the process $pp\to H^\pm H^\pm jj$ is that its cross section is approximately  proportional to the square of the mass splitting $\Delta m$ \cite{Arhrib:2019ywg}. The dependence of the cross sections $\sigma(pp\to H^\pm H^\pm jj)$  for different mass splitting $\Delta m$ is depicted in Fig.~\ref{Fig:CS}. During the calculation, {\bf Madgraph5\_aMC@NLO} \cite{Alwall:2014hca} is employed with the  preselection cuts for VBF processes at parton level 
\begin{equation}\label{Eqn:Cut1}
\eta_{j_1}\times \eta_{j_2} <0,~|\Delta \eta_{jj}|>2.5.
\end{equation}
From Fig.\ref{Fig:CS}, it can be seen that following the enlargement of $ \Delta m $ from $\Delta m = 100$ GeV to $\Delta m = 300$ GeV, the production cross section is enlarged by about ten times for the same value of $m_{H^\pm}$. In the actual model, $m_A\lesssim m_{H^\pm}$ should be satisfied. Thus, 
the results of BPs in Table \ref{Tab:BP} are further illustrated. It is obvious that at the 14 TeV HL-LHC, the cross section usually increases as $m_{H^\pm}$ becomes larger when $m_{H^\pm}\lesssim 250$ GeV. While for  $m_{H^\pm}\gtrsim 250$ GeV, the cross section does not change so much as $m_{H^\pm}$ increases. At the 27 TeV HE-LHC, the cross section always tends to increase when $m_{H^\pm}$ increases, which is about three to four times larger than it at the 14 TeV HL-LHC.

Now we discuss the same-sign dilepton signature and its corresponding backgrounds at hadron colliders. The full process of such signature is
\begin{equation}
	pp\to W^{\pm *}W^{\pm *}j j \to H^\pm H^\pm jj \to (W^\pm H) (W^\pm H)jj \to (l^\pm \nu)H (l^\pm \nu)H jj \to l^\pm l^\pm \cancel{E}_Tjj,
\end{equation}
in the IDM, where $j$ is the forward and energetic jet from the initial parton, and the leptons contain electron and muon ($l=e,\mu$). In the following, we choose BP10, BP15, and BP20 in Table \ref{Tab:BP} to show the distribution of certain variables and corresponding cut flow at colliders.

The main SM backgrounds come from $W^\pm W^\pm jj$, $WZjj$,  $ZZjj$, $VVVjj$, and $t\bar{t}V$.
Both the strong and electroweak production of the $VVjj$ process are taken into account. According to the experimental result of ATLAS collaboration \cite{Aaboud:2019nmv}, there should be additional contributions from $V\gamma$, electron charge misreconstruction, and non-prompt leptons, which are sub-dominant and thus are not taken into account in this work. After generating the parton level events for all BPs and corresponding SM backgrounds using {\bf Madgraph5\_aMC@NLO} \cite{Alwall:2014hca}, {\bf Pythia8} \cite{Sjostrand:2007gs} is used for parton showering and hadronization. Finally, the detector simulation is performed with {\bf Delphes3} \cite{deFavereau:2013fsa}. In this work, all the signals and backgrounds are simulated at the leading order.

After the above simulation, several cuts are applied to highlight the signal, which are simply categorized into four parts, i.e, cuts-1 to cuts-4. First, cuts-1 aims to select the same-sign dilepton signature, where we require exactly two leptons carrying the same charge in the final states,
\begin{equation}\label{Eqn:Cut2}
N(l^\pm) = 2, P_T^{l^\pm}>20~\text{GeV}, |\eta_{l^\pm}|<2.5,
\end{equation}

Then in cuts-2 for the forward jet pair, events with at least two jets and with $b$-jet veto can pass the selection
\begin{equation}\label{Eqn:Cut3}
N(j) \geq 2, P_T^j>30~\text{GeV},|\eta_j|<5,N(b)=0.
\end{equation}
Here, the $b$-jet veto criteria is to suppress the $t\bar{t}V$ background. As shown in Table ~\ref{Tab:cut14}, at this level of cuts, the SM background is about three orders of magnitudes larger than the signal. Therefore, additional cuts are expected to further eliminate the background.

\begin{figure}[!htbp]
	\begin{center}
		\includegraphics[width=0.46\linewidth]{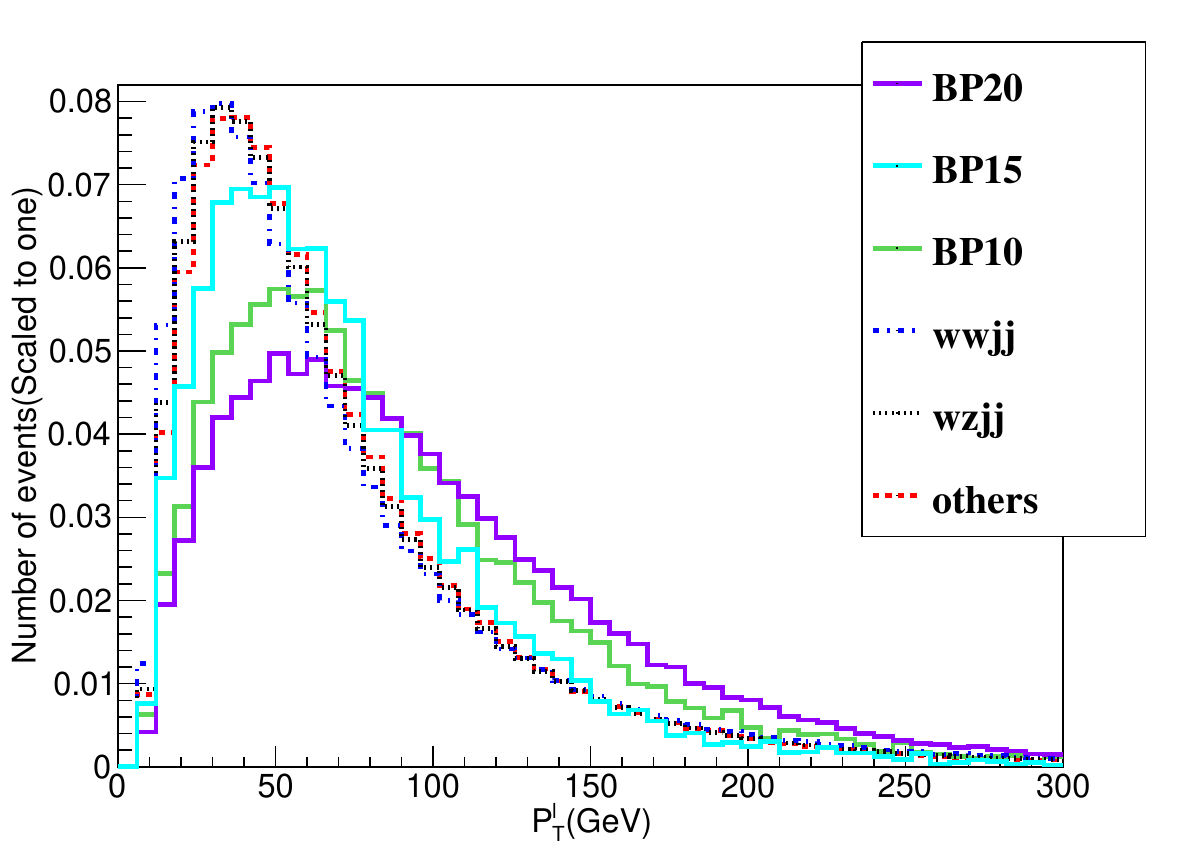}
		\includegraphics[width=0.46\linewidth]{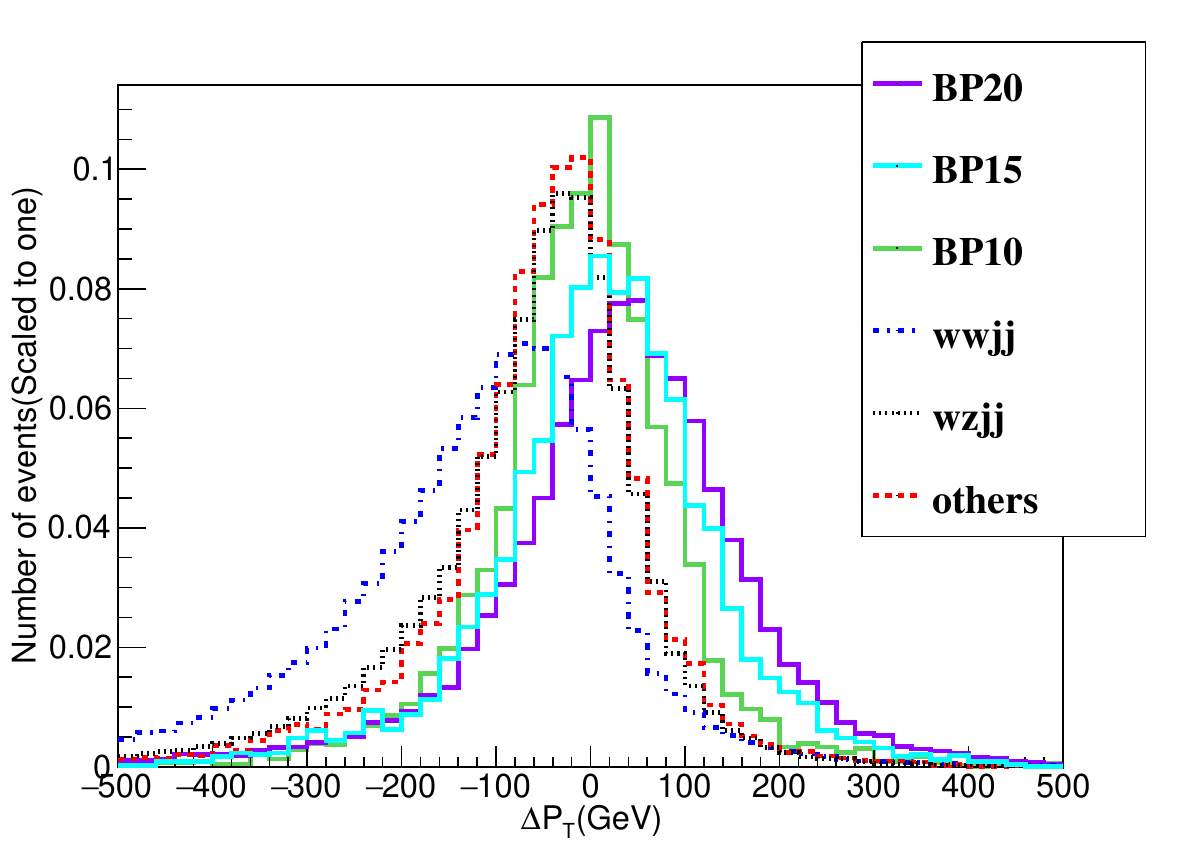}
		\\
		\includegraphics[width=0.46\linewidth]{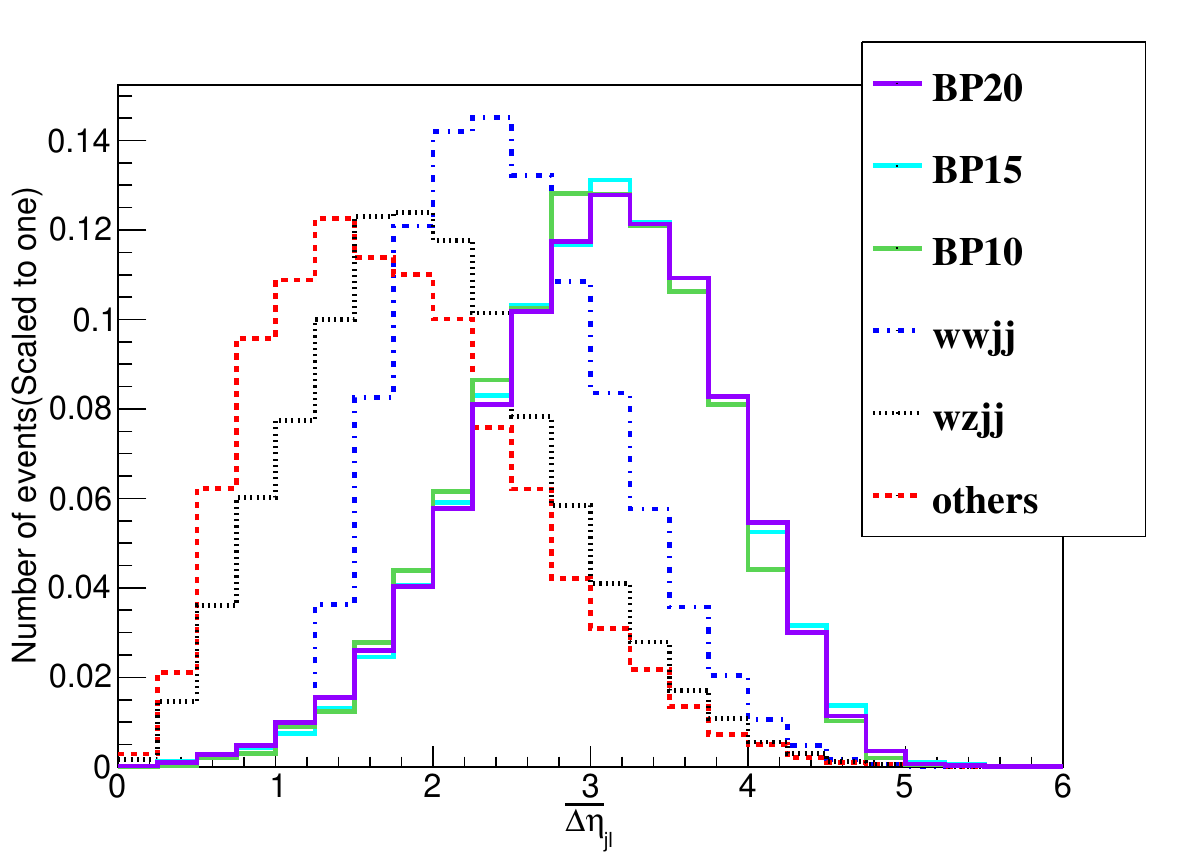}
		\includegraphics[width=0.46\linewidth]{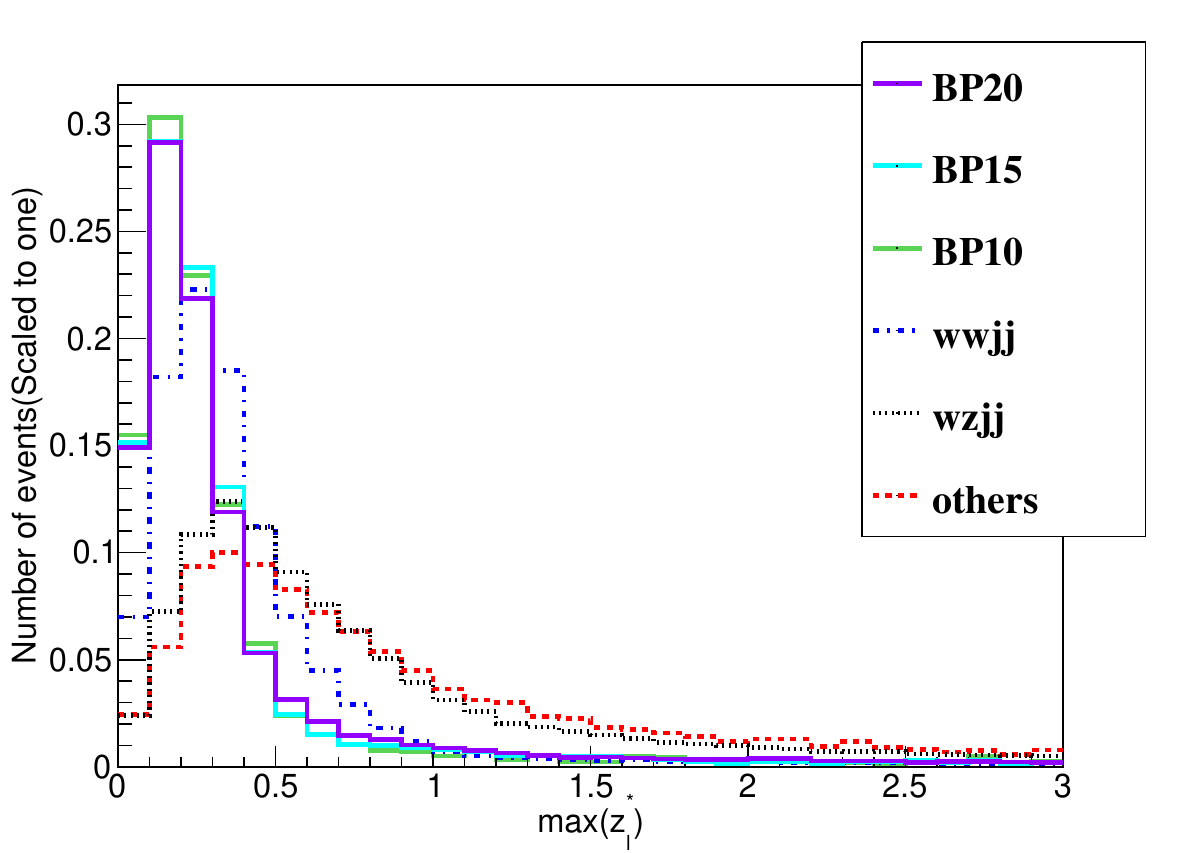}
		\\
		\includegraphics[width=0.46\linewidth]{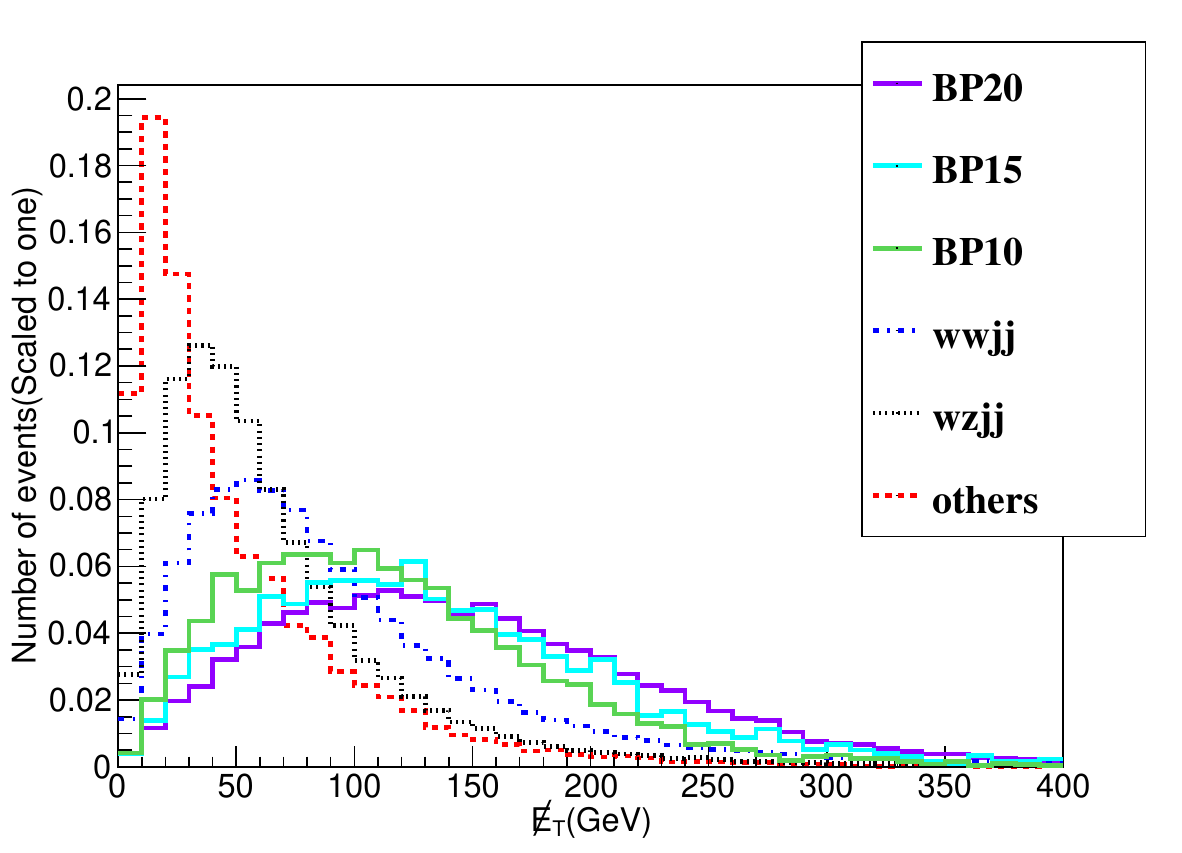}
		\includegraphics[width=0.46\linewidth]{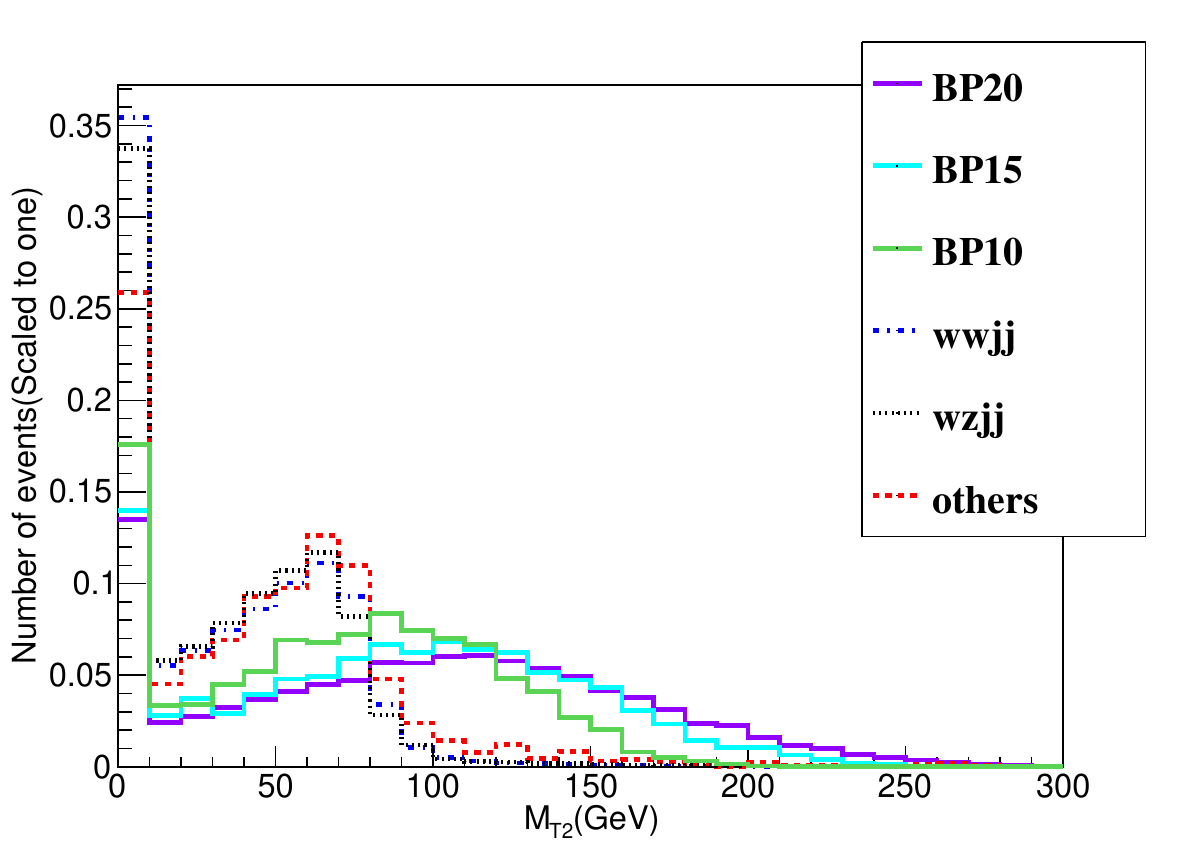}
	\end{center}
	\caption{Normalized distribution of $P_T^l$ (up-left panel), $\Delta P_T$ (up-right panel),  $\overline{\Delta \eta}_{jl}$ (middle-left panel), max$(z_{l}^{*})$ (middle-right panel), $\cancel{E}_T$ (down-left panel), and $M_{T2}$ (down-right panel) variable for BP10, BP15, BP20 (solid line) and corresponding SM backgrounds (dashed line) at $\sqrt{s} = 14$ TeV. The $P_T^l$, $\Delta P_T$, $\overline{\Delta \eta}_{jl}$, max$(z_{l}^{*})$, and $\cancel{E}_T$ variables are drawing after the cuts in Eqn.\eqref{Eqn:Cut1},\eqref{Eqn:Cut2},\eqref{Eqn:Cut3} are applied, while $M_{T2}$ variable is drawing after cuts in Eqn.\eqref{Eqn:Cut1},\eqref{Eqn:Cut2},\eqref{Eqn:Cut3} and  $\cancel{E}_T > 100$ GeV are applied.}
	\label{Fig:Dis14}
\end{figure}

To seek for proper cut criteria, the normalized distribution of certain parameters is shown in Fig.~\ref{Fig:Dis14}. Specifically speaking, the up-left panel shows the  $P_T^l$ variable. The distributions of $P_T^l$ are not well separated for signal and background. Instead, we consider the $\Delta P_T$ parameter, defined as $\Delta P_T=(P_{T}^{l_{1}}+P_{T}^{l_{2}}) - (P_{T}^{j_{1}}+P_{T}^{j_{2}})$, which is shown in the up-right panel. For the BP signals, the distributions of  $\Delta P_T$ tend to be larger than those of the backgrounds. Based on this feature, we require $\Delta P_T>0$. That is to say, the scalar sum of the transverse momentum of two leptons is larger than the scalar sum of the transverse momentum  of  leading and sub-leading jet. In the middle-left panel, we depict the distribution of $\overline{\Delta \eta}_{jl}$ variable, where $\overline{\Delta \eta}_{jl}$ is defined as
\begin{equation}
\overline{\Delta \eta}_{jl} = \sqrt{ \sum_{m=1}^{2} \sum_{n=1}^{2}\frac{(\eta_{jm} - \eta_{ln})^{2}}{4}}.
\end{equation}
Here, $\eta_{jm}$ are the pseudorapidity of leading and sub-leading jets with $m = 1, 2$ and $\eta_{ln}$ are the pseudorapidity of leading and sub-leading leptons with $n = 1, 2$. The $\overline{\Delta \eta}_{jl}$ variable characterizes the averaged pseudorapidity separation between jets and leptons, where $ \overline{\Delta \eta}_{jl} $ variable larger than three is good enough to separate the signal and background. Another distinguishable variable used by the experimental groups is the Zeppenfeld variable $z_{l}^{*}$, which is defined as \cite{Rainwater:1996ud}  
\begin{equation}
z_{l}^{*} = ~\left|\eta_{l} - \frac{\eta_{j1} + \eta_{j2}}{2}\right| / |\eta_{j1} - \eta_{j2}|.
\end{equation}
The max$(z_l^*)$ variable is used by CMS collaboration to define the $W^\pm W^\pm jj$ and $WZjj$ signal region \cite{Sirunyan:2020gyx}. Since both $W^\pm W^\pm jj$ and $WZ jj$ are the backgrounds in this work, we use a more stringent cut max$(z_l^*)<0.3$, i.e., the largest $ z_{l}^{*} $ variable less than 0.3. In summary, the cuts-3 we adopted are 
\begin{equation}\label{Eqn:Cut4}
\Delta P_T>0 , ~\overline{\Delta \eta}_{jl} > 3, ~\text{max}(z_{l}^{*})< 0.3. 
\end{equation}

\begin{table}
	\begin{center}
		\begin{tabular}{c  l l l l l l } 
			\hline
			\hline
			\textbf{Cross section (fb)}~ & \textbf{BP10} & \textbf{BP15} & \textbf{BP20} & $W^{\pm} W^{\pm}jj$ & $WZjj$ & Others \\
			\hline
			Preselection & $1.88\times10^{-2}$~ & $1.89\times10^{-2}$~ & $2.04\times10^{-2}$~ & $1.35\times10^{1}$~ &$5.50\times10^{1}$~ & $3.05\times10^{0}$ \\
			\hline
			$N(l^\pm) = 2,P_T^{l^\pm}>20~\text{GeV}$ \\
			$|\eta_{l^\pm}|<2.5$ & $1.01\times10^{-2}$ & $1.08\times10^{-2}$ & $1.19\times10^{-2}$& $5.29\times10^{0}$ & $6.43\times10^{0}$ & $3.66\times10^{-1}$\\
			\hline
			$N(j) \geq 2, P_T^j>30$ GeV\\ $|\eta_j|<5,N(b)=0$ & $8.62\times10^{-3}$ & $9.13\times10^{-3}$  & $1.21\times10^{-2}$  & $4.60\times10^{0}$ & $5.43\times10^{0}$ & $2.05\times10^{-1}$\\
			\hline
			$\Delta P_T>0,\overline{\Delta {\eta}}_{jl} > 3$  \\ max$(z_{l}^{*})< 0.3$   & $1.56\times10^{-3}$  & $2.48\times10^{-3}$  & $3.06\times10^{-3}$  & $1.34\times10^{-1}$  & $2.834\times10^{-2}$  & $1.12\times10^{-3}$ \\
			\hline
			$\cancel{E}_T>100$ GeV \\ $M_{T2}>100$ GeV & $3.71\times10^{-4}$  & $8.41\times10^{-4}$  & $1.33\times10^{-3}$  & $7.31\times10^{-4}$  & $1.10\times10^{-4}$  & $8.87\times10^{-5}$  \\
			\hline
			\hline
			Significance  & \quad 0.67 & \quad 1.52 & \quad 2.39 & \quad--- & \quad---& \quad---\\
			\hline
		\end{tabular}
	\end{center}
	\caption{Cut flow table for BP10, BP15, BP20 signal and various background process at $\sqrt{s}=14$ TeV. The $ZZjj$, $VVVjj$ and $t\bar{t}V$ backgrounds are classified as others for their contributions to the total backgrounds are blow $10\%$ after applying all cuts. The significance $ S/\sqrt{B} $ is calculated by assuming an integrated luminosity $\mathcal{L} = 3~\text{ab}^{-1}$.
		\label{Tab:cut14}}
\end{table}

The results for both signal and background at the level of cuts-3 are shown in the fourth row
of Table ~\ref{Tab:cut14}. At this level, the cross section of the $ZZjj$, $VVVjj$ and $t\bar{t}V$ backgrounds are smaller than the signal. The dominant ones are  $W^\pm W^\pm jj$ and $WZ jj$. From Fig.~\ref{Fig:Dis14}, it is also clear that for the $W^{\pm} W^{\pm} jj$ background process, $\Delta P_T$, $\overline{\Delta \eta}_{jl}$, and $z_{l}^{*}$ variables are not so distinguishable from signal to background. This is because the main part of $W^\pm W^\pm jj$ generated from electroweak production process has a similar topological structure to signal.

In order to suppress the $W^\pm W^\pm jj$ background efficiently, more advanced cuts should be applied.
Despite  the additional two forward jets in the same-sign dilepton signature, the decay chain of charged scalar $H^\pm \to W^\pm H\to l^\pm \nu H$ is actually the same as the decay chain of chargino $\tilde{\chi}_1^\pm \to W^\pm \tilde{\chi}_1^0\to l^\pm \nu \tilde{\chi}_1^0$, which means we can apply similar cuts for opposite-sign dilepton signature as in Ref. \cite{Aad:2019vnb}. Here, we take the variables $\cancel{E}_T$ and $M_{T2}$ into account.
The $M_{T2}$ variable is defined as \cite{Lester:1999tx,Barr:2003rg},
\begin{equation}
M_{T2}  = \underset{\textbf{q}_{T,1} + \textbf{q}_{T,2} =~ \cancel{\textbf{E}}_T}{\text{min}} \left\{\text{max}\left[M_{T} (\textbf{P}_{T}^{l_{1}},\textbf{q}_{T,1}),M_{T}(\textbf{P}_{T}^{l_{2}},\textbf{q}_{T,2})\right]\right\},
\end{equation}
where $\textbf{P}_{T}^{l_{1}}$ and $\textbf{P}_{T}^{l_{2}}$ are the transverse momentum vectors of the two leptons, $\textbf{q}_{T,1}$ and $\textbf{q}_{T,2}$ are all possible combinations of two transverse momentum vectors that satisfy $\textbf{q}_{T,1} + \textbf{q}_{T,2} =\cancel{\textbf{E}}_T$. The $M_{T2}$ variable is calculated by applying the algorithms proposed in Ref.~\cite{Lester:2014yga}. 

Distributions of $\cancel{E}_T$ and $M_{T2}$ are also shown in Fig.\ref{Fig:Dis14}. For the signal process, both neutrinos $\nu$ and dark matter $H$ contribute to the missing transverse energy $\cancel{E}_T$, which thus usually leads to a larger $\cancel{E}_T$ than the backgrounds. It also can be seen that the $M_{T2}$ variable serves as the most efficient cut. Because theoretically, this variable can not exceed the mass of $W$ boson at $ m_W = 80.4 $ GeV for the background process, while the theoretical upper limit for the signal process is the mass of $ m_{H^{\pm}} $. Therefore, when exceeding $80$ GeV, the $ M_{T2} $ variable decreases severely and nearly vanishing when $M_{T_2}>100$ GeV for all background processes. But for the signals, it still have a large part existed, especially for BP20 due to the largest $ M_{H^{\pm}} $ it possessed. 
In short, we require that missing transverse energy $\cancel{E}_T$ is greater than $ 100 $ GeV and $M_{T2}$ variable is greater than $ 100$ GeV for cuts-4:
\begin{equation}\label{Eqn:Cut5}
\cancel{E}_T > 100 ~\text{GeV}, ~M_{T2} > 100~ \text{GeV}.
\end{equation}

Results after applying cuts-4 are shown in the fifth row of Table ~\ref{Tab:cut14}. After applying all of the cuts, the contributions of $ W^{\pm} W^{\pm} jj$ process to the total backgrounds is great than $80\%$, the contributions of $ WZjj $ process to the total backgrounds is great than $ 10\% $ and the part named as others are comes from the sum of $ t \bar{t} V$, $VVVjj$, $ZZjj$ process, which serves less than $10\%$ contributions to the total backgrounds. For the signal process after the full cuts are applied, BP20 has the largest cross section, because it both possesses the largest $ \Delta m $ and $ m_{H^{\pm}} $, the former leads to a large cross section in simulation and the latter leads to the highest efficiency when passing the cut flow. The total cross section of backgrounds is $9.30\times10^{-4}$~fb, which is larger than it of BP10 and BP15, but is smaller than it of BP20. Although a good signal-to-background ratio is achieved, the cross sections for signals after all cuts are a little bit too small to probe. For instance, assuming an integrated luminosity $\mathcal{L} = 3~\text{ab}^{-1}$, we expect about 1.1 events for BP10, 2.5 events for BP15 and 4.0 events for BP20 with 2.8 events for total backgrounds, of which the corresponding significance $S/\sqrt{B}$ is 0.67, 1.52 and 2.39, respectively. Therefore, the same-sign dilepton signature is not so promising at the HL-LHC.

Before ending the discussion on the 14 TeV simulation, let's briefly summarize the searching strategy.
The main backgrounds come from $ W^{\pm} W^{\pm} jj$ and $ WZjj$ process with the  production cross section over 10~fb after preselection cuts. Due to the similar distributions of certain variables (such as $P_T^{l^\pm},z_l^*$) from the dominant background to the signal, we can only choose the simplest cuts in cuts-1 and cuts-2. Then, we have to apply the cuts extremely in cuts-3 and cuts-4, even if this shall lead to a faint signal and low significance. The above analysis can be improved by considering more sophisticated selection criteria, such as employing a boosted decision tree, which is beyond the scope of this work. Instead, we further consider the same-sign dilepton signature at the 27 TeV HE-LHC.

\begin{figure}[!htbp]
	\begin{center}
		\includegraphics[width=0.46\linewidth]{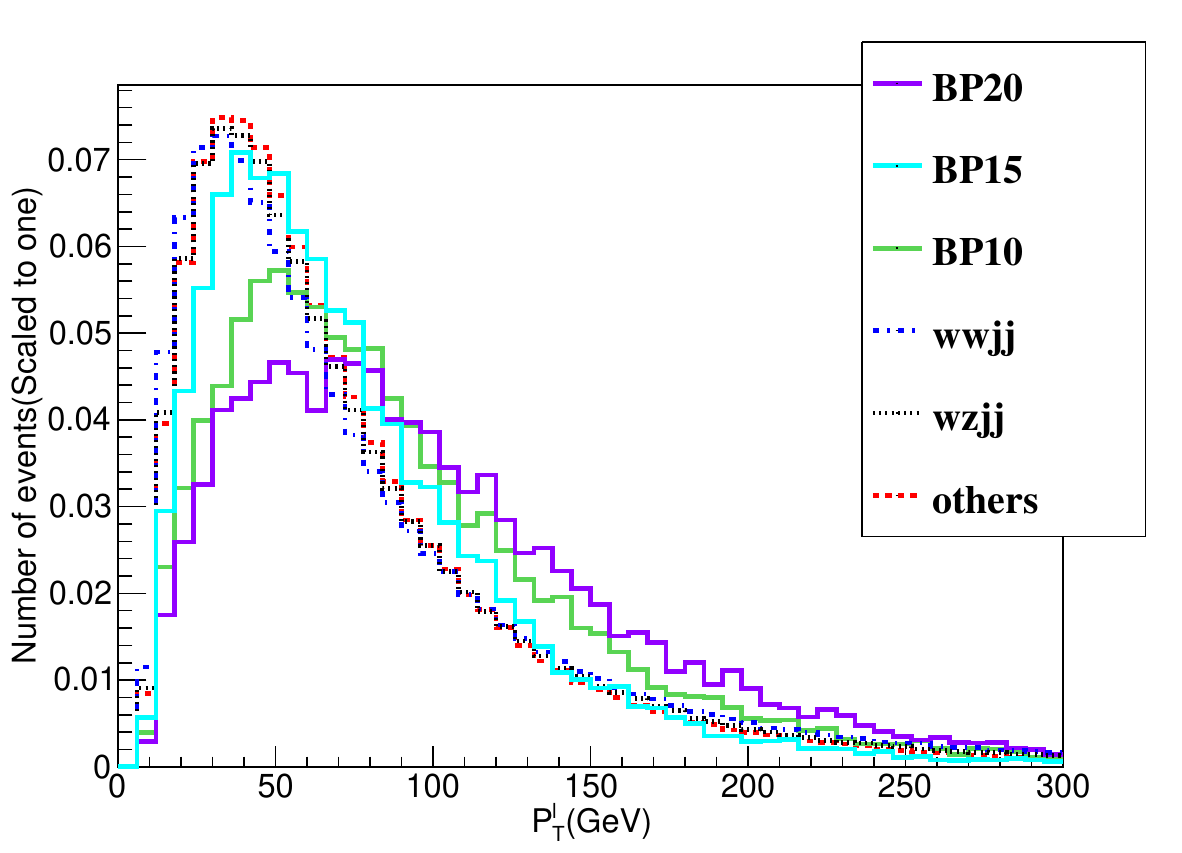}
		\includegraphics[width=0.46\linewidth]{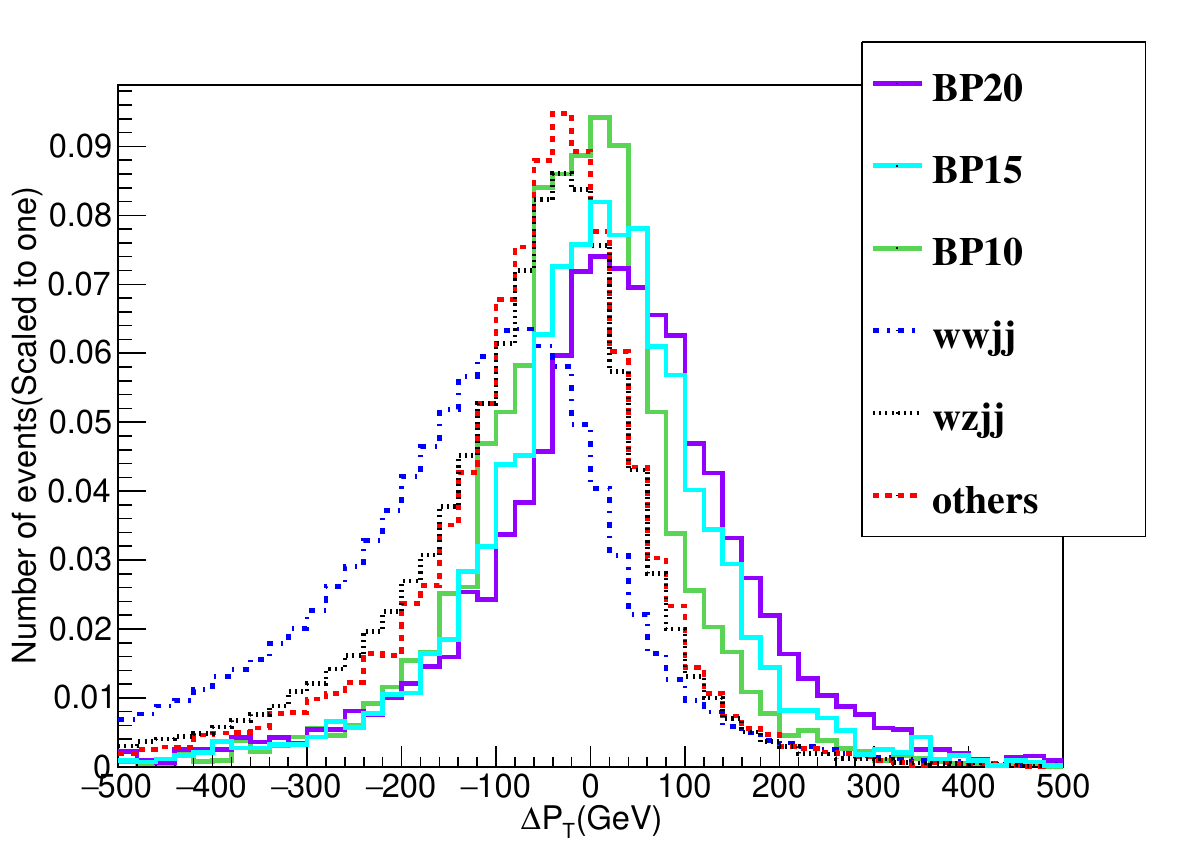}
		\\
		\includegraphics[width=0.46\linewidth]{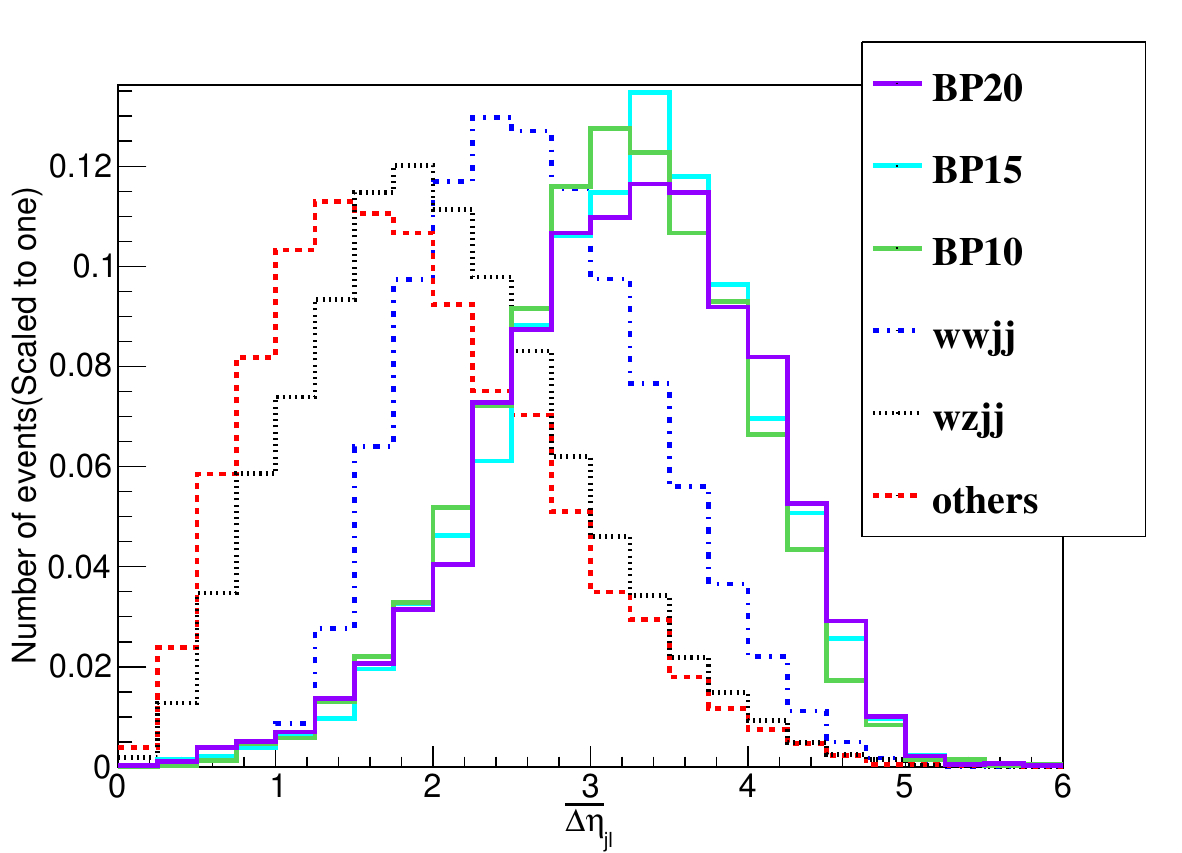}
		\includegraphics[width=0.46\linewidth]{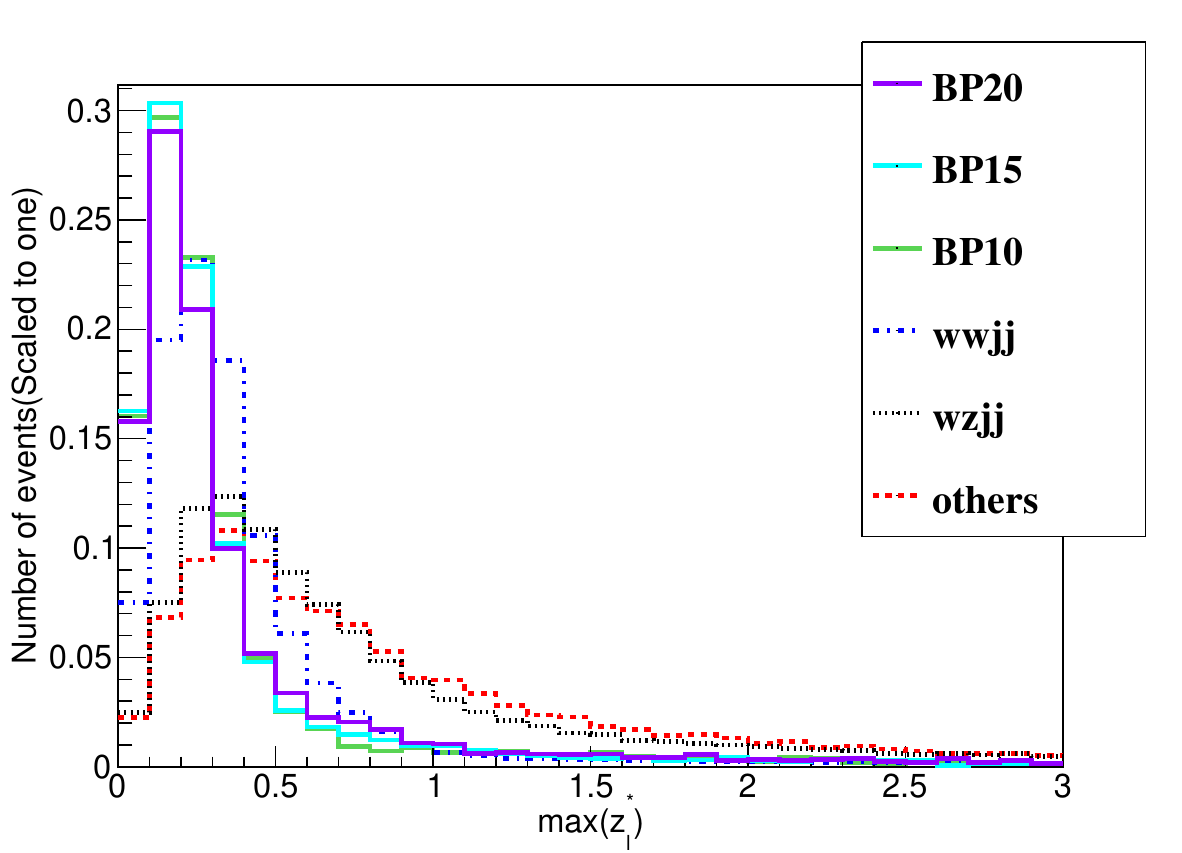}
		\\
		\includegraphics[width=0.46\linewidth]{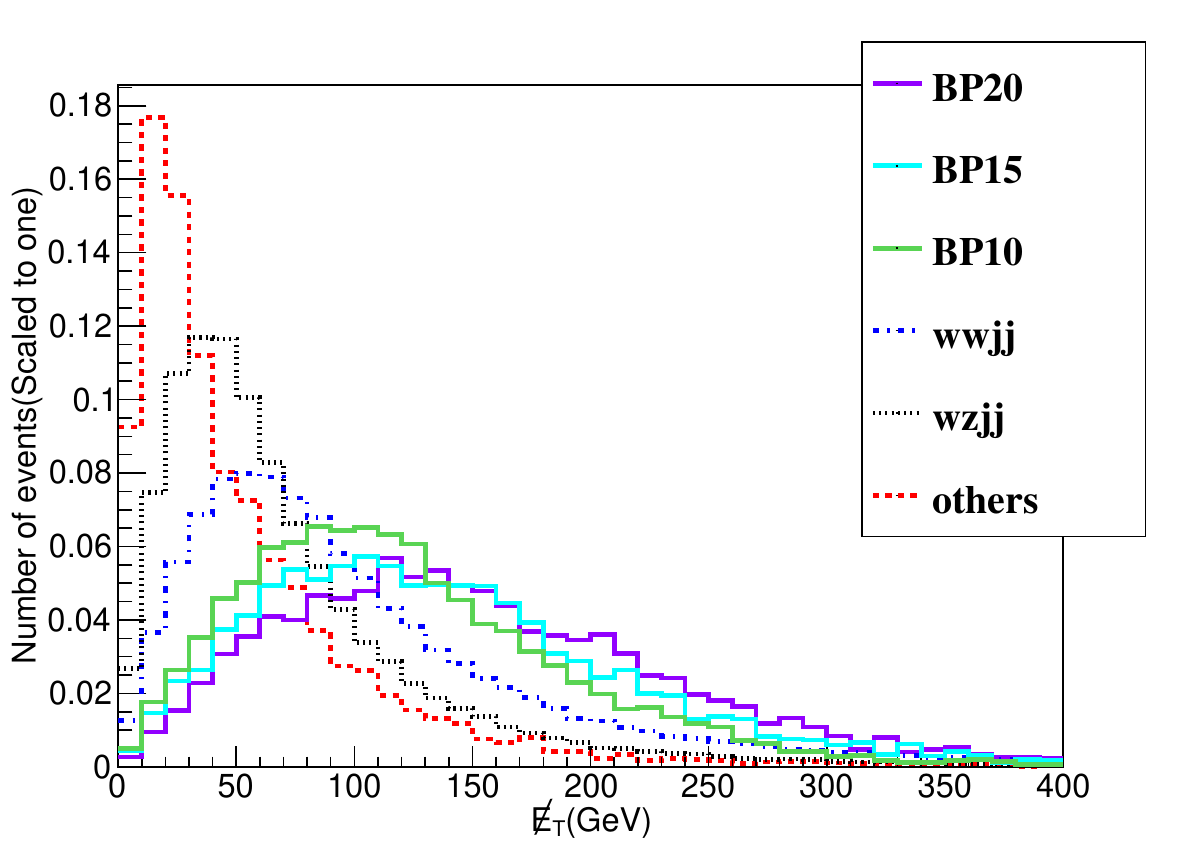}
		\includegraphics[width=0.46\linewidth]{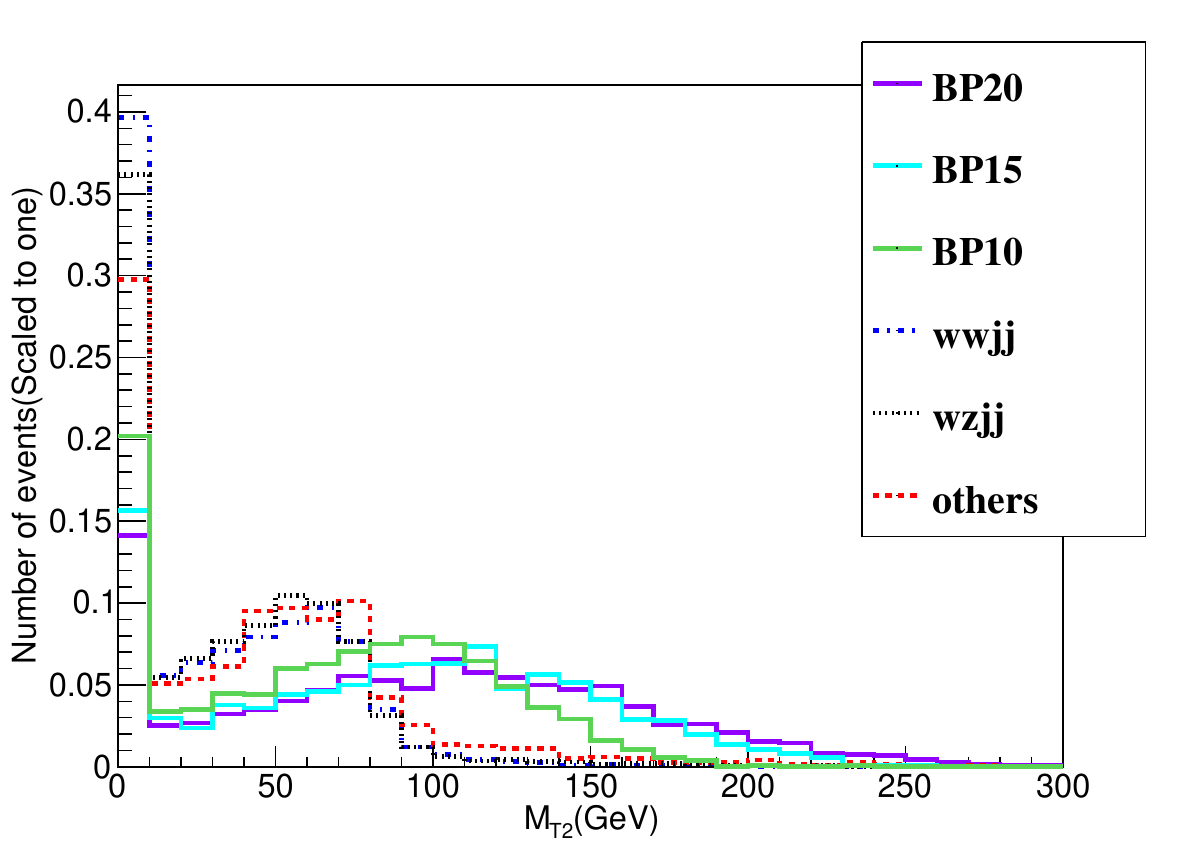}
	\end{center}
	\caption{Same as Fig.\ref{Fig:Dis14}, but at $\sqrt{s}=27$ TeV}
	\label{Fig:Dis27}
\end{figure}

The normalized distribution of $P_T^l$, $\Delta P_T$, $\overline{ \Delta \eta}_{jl} $, max$(z_{l}^{*})$, $ \cancel{E}_T$ and $ M_{T2} $ variable at 27 TeV are shown in Fig.~\ref{Fig:Dis27}, which are similar to the results of 14 TeV. Hence, we adopt the same criteria as 14 TeV for cuts-1 to cuts-3. Meanwhile, considering the fact that the final states of neutrinos $\nu$ as well as dark matter $H$ are more energetic at 27 TeV than those at 14 TeV, we slightly tighten cuts-4 as
\begin{equation}\label{Eqn:Cut7}
\cancel{E}_T>110~\text{GeV}, ~M_{T2}>125~\text{GeV}.
\end{equation}
The cross section for both signal and backgrounds with the cut flow are listed in Table.\ref{Tab:cut27}. After applying the full cuts, at $\sqrt{s}=27$ TeV only two processes have considerable contributions to the total backgrounds. The main part of total backgrounds comes from $W^{\pm} W^{\pm} jj$ process, with the contribution to the total backgrounds great than $85\%$. The rest part of the total backgrounds comes from $WZjj$ process. The part named as others comes from the sum of $ZZjj$, $VVVjj$, and $t\bar{t}V$ process has negligible contributions to the total backgrounds due to the more stringent cuts we have used. With larger production cross section and higher luminosity $\mathcal{L} = 15~\text{ab}^{-1}$, we find that the dilepton signature is promising for some benchmark points at the 27 TeV HE-LHC. Quantitatively speaking , we expect about 8 events for BP10, 31 events for BP15 and 56 events for BP20 with 48 events for total backgrounds, of which the corresponding significance $S/\sqrt{B}$ is 1.21, 4.54, and 8.08, respectively. 

\begin{table}
	\begin{center}
		\begin{tabular}{c l l l l l l} 
			\hline
			\hline
			\textbf{Cross section(fb)}~ & \textbf{BP10} & \textbf{BP15} & \textbf{BP20} & $W^{\pm} W^{\pm}jj$ & $WZjj$ & Others \\
			\hline
			Preselection & $6.57\times10^{-2}$~  & $7.40\times10^{-2}$~  & $8.59\times10^{-2}$~  & $4.61\times10^{1}$~ & $1.95\times10^{2}$~ & $1.38\times10^{1}$ \\
			\hline
			$N(l^\pm) = 2,P_T^{l^\pm}>20~\text{GeV}$ \\
			$|\eta_{l^\pm}|<2.5$   & $3.19\times10^{-2}$  & $3.77\times10^{-2}$  & $4.54\times10^{-2}$  & $1.47\times10^{1}$ & $1.95\times10^{1}$ & $1.32\times10^{0}$\\
			\hline
			$N(j) \geq 2, P_T^j>30$ GeV\\ $|\eta_j|<5,N(b)=0$  & $2.49\times10^{-2}$ & $3.06\times10^{-2}$ & $3.74\times10^{-2}$  & $1.13\times10^{1}$ & $1.58\times10^{1}$ & $8.67\times10^{-1}$ \\
			\hline
			$\Delta P_T>0,\overline{\Delta {\eta}}_{jl} > 3$  \\ max$(z_{l}^{*})< 0.3$  & $6.13\times10^{-3}$  & $9.74\times10^{-3}$  & $1.23\times10^{-2}$  & $4.94\times10^{-1}$  & $1.21\times10^{-1}$  & $4.33\times10^{-3}$  \\
			\hline
			$\cancel{E}_T>110$ GeV \\ $M_{T2}>125$ GeV  & $5.58\times10^{-4}$  & $2.09\times10^{-3}$  & $3.72\times10^{-3}$  & $2.79\times10^{-3}$  & $3.90\times10^{-4}$  & \quad0\\
			\hline
			\hline
			Significance & \quad 1.21 & \quad 4.54 & \quad 8.08 & \quad---& \quad---&\quad---\\
			\hline
		\end{tabular}
	\end{center}
	\caption{Cut flow table for BP10, BP15, BP20 signal and various background process at $\sqrt{s}=27$ TeV. The $ZZjj$, $VVVjj$ and $t\bar{t}V$ backgrounds are classified as others for it's contributions to the total backgrounds are negligible after applying all cuts. The significance $ S/\sqrt{B} $ is calculated by assuming an integrated luminosity $\mathcal{L} = 15~\text{ab}^{-1}$.
		\label{Tab:cut27}}
\end{table} 

\begin{figure}
	\begin{center}
		\includegraphics[width=0.8\linewidth]{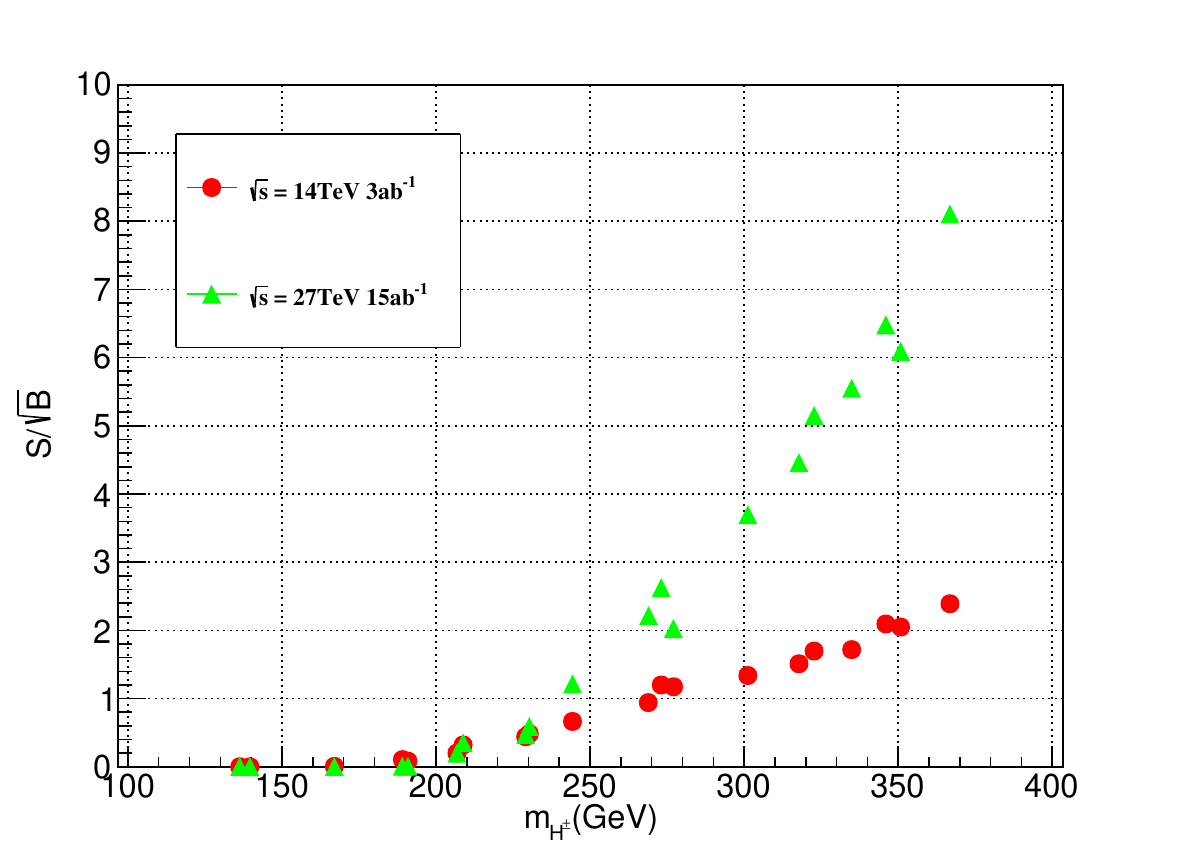}
	\end{center}
	\caption{Significance of all twenty BPs at  $\sqrt{s} = 14$ TeV, $\mathcal{L} = 3~\text{ab}^{-1}$(red points) and $\sqrt{s} = 27$ TeV, $\mathcal{L} = 15~\text{ab}^{-1}$(green points).}
	\label{Fig:Sig}
\end{figure}

Finally, based on the cuts adopted in the above discussion, we extend our analysis to all the twenty benchmark points listed in Table ~\ref{Tab:BP}. The results of significance for BPs at 14 TeV HL-LHC and 27 TeV HE-LHC are shown in Fig.\ref{Fig:Sig}. It can be seen that following the increase of $m_{H^{\pm}}$, the significance increased, for a larger $ m_{H^{\pm}} $ leading to a higher cut efficiency. At $\sqrt{s} = 14$ TeV limited by the faint signal, even for BP20 with the largest $m_{H^{\pm}}$, the significance can only slightly excess two. At $\sqrt{s} = 27$ TeV, with a larger cross section and higher luminosity, we find that BP16 to BP20 can have a significance larger than five.
That is to say the promising region of same-sign dilepton signature at $\sqrt{s}=27$ TeV is $250~\text{GeV}\lesssim m_{H^\pm}-m_H\lesssim 300$ GeV with dark matter mass $m_H\sim 60$ or 71 GeV. 

\section{Conclusion}\label{SEC:CL}
The IDM is a 2HDM imposed with an exact $ Z_{2} $ symmetry, which leads to a DM candidate. This model gives rise to rich phenomenology, which has been extensively studied. In this paper, we perform a detailed analysis on the same-sign dilepton signature $pp\to W^{\pm *}W^{\pm *}jj \to H^\pm H^\pm jj\to (l^\pm \nu)H (l^\pm \nu)H jj \to l^\pm l^\pm \cancel{E}_Tjj$ in the IDM, where $H$ is the DM candidate. According to our simulation, this signature is promising for large mass splitting $\Delta m= m_A-m_H$, which is complementary to the well studied opposite-sign dilepton signature. 

We first perform a random scan over the low mass region of IDM with various constraints taken into account. By requiring the relic density within $3\sigma$ range of the Planck observation value $\Omega h^2 = 0.1200\pm 0.0012$, we find three viable parameter space. One is the Higgs resonance region around $m_H\lesssim m_h/2$. Another one is the vector boson annihilation region around $m_H\sim71.5$ GeV. The third one is the coannihilation region with $m_A-m_H\sim 8$ GeV and $m_H\sim 65$ GeV. Since the coannihilation region gives a vanishing cross section of same-sign dilepton signature, we select twenty benchmark points from the Higgs resonance and  vector boson annihilation region, which are listed in Table \ref{Tab:BP}.

We then simulate the same-sign dilepton signature  for the BPs as well as SM backgrounds both at $\sqrt{s} = 14$ TeV HL-LHC and $\sqrt{s} = 27$ TeV HE-LHC. With similar decay topological structure to signal, the dominant background comes from  $W^{\pm} W^{\pm} jj$. The most efficient cut to suppress background is $M_{T2}$ variable. According to our simulation, at $\sqrt{s} = 14$ TeV with luminosity $\mathcal{L} = 3\text{ab}^{-1}$, at best four signal events can survived  after applying the full cuts. Limited by the number of signal events, the BPs can only achieve a significance slightly larger than two. At $\sqrt{s} = 27$ TeV with luminosity $\mathcal{L} = 15\text{ab}^{-1}$, we can probe benchmark points with large mass $\Delta m$. For example, BP16 to BP20 can have a significance larger than five.

In a nut shell, the same-sign dilepton signature is not promising at $\sqrt{s}=14$ TeV HL-LHC, but is promising at $\sqrt{s}=27$ TeV with the viable region of $250~\text{GeV}\lesssim m_{H^\pm}-m_H\lesssim 300$ GeV and dark matter mass $m_H\sim 60$ or 71 GeV.

\section*{Acknowledgments}

This work is supported by the National Natural Science Foundation of China under Grant No. 11805081, Natural Science Foundation of Shandong Province under Grant No. ZR2019QA021 and ZR2018MA047.


\end{document}